\documentclass[a4paper,aps,prd,reprint,nofootinbib,notitlepage,onecolumn]{revtex4-1}

\usepackage{amsmath}
\usepackage{amssymb}
\usepackage{dcolumn}
\usepackage{bm}
\usepackage{color}
\usepackage{dsfont}
\usepackage{epstopdf}
\usepackage{graphicx}
\usepackage{soul}
\usepackage{enumerate}

\newcommand{\bea}{\begin{eqnarray}}
\newcommand{\eea}{\end{eqnarray}}
\newcommand{\beq}{\begin{eqnarray}}
\newcommand{\eeq}{\end{eqnarray}}

\newcommand{\be}{\begin{equation}}
\newcommand{\ee}{\end{equation}}

\begin{document}
\allowdisplaybreaks

\title{Pad{\'e} approximants and analytic continuation of Euclidean $\Phi$-derivable approximations}

\author{Gergely Mark{\'o}}
\email{smarkovics@hotmail.com}
\affiliation{Institute of Physics, E{\"o}tv{\"o}s University, P\'azm\'any P{\'e}ter s{\'e}t\'any 1/A, H-1117 Budapest, Hungary}

\author{Urko Reinosa}
\email{reinosa@cpht.polytechnique.fr}
\affiliation{Centre de Physique Th{\'e}orique, {\'E}cole Polytechnique, CNRS, Universit{\'e} Paris-Saclay, F-91128 Palaiseau, France}

\author{Zsolt Sz{\'e}p}
\email{szepzs@achilles.elte.hu}
\affiliation{MTA-ELTE Statistical and Biological Physics Research Group, H-1117 Budapest, Hungary}

\begin{abstract}
We investigate the Pad{\'e} approximation method for the analytic continuation of numerical data and its ability to access, from the Euclidean propagator, both the spectral function and part of the physical information hidden in the second Riemann sheet. We test this method using various benchmarks at zero temperature: a simple perturbative approximation as well as the two-loop $\Phi$-derivable approximation. The analytic continuation method is then applied to Euclidean data previously obtained in the $O(4)$ symmetric model (within a given renormalization scheme) to assess the difference between zero-momentum and pole masses, which is in general a difficult question to answer within nonperturbative approaches such as the $\Phi$-derivable expansion scheme.
\end{abstract}

\maketitle

\section{Introduction}

In recent years, much progress was achieved in accessing the properties of interacting quantum fields in equilibrium. Part of the success stems from the development of continuum nonperturbative tools \cite{Braun:2007bx,Braun:2014ata,Contant:2017gtz,Reinhardt:2017pyr} that allow to implement some of the resummations of perturbative diagrams that are required in this context. Most of these approaches are formulated in Euclidean space (or imaginary-time formalism) because, on general grounds, the equations are easier to solve than the corresponding ones in Minkowski space (or real-time formalism). For instance, to date, there is no complete Minkowski space study of the temperature driven symmetry restoration in the four dimensional $O(N)$ model which could compete in precision with its Euclidean counterpart. Yet, certain quantities of interest, such as transport coefficients, pole masses and decay rates, require the use of the real-time formulation, and, therefore, a constant effort is put into extending the previous methods (not only formally but also at a computational level) to Minkowski space. In particular, the self-consistent scalar propagator equation was solved in Minkowski space at zero and finite temperatures, both in the symmetric and broken phases (see e.g. \cite{Sauli:2001mb, VanHees:2001pf, Cooper:2004rs, Sauli:2004bx, Arrizabalaga:2007zz, Jakovac:2006dj, Jakovac:2006gi}). These spectral function based numerical solutions were obtained in various  nonperturbative approximations. More recently, the spectral function was accessed using an analytic continuation of the Functional Renormalization Group (FRG) equations \cite{Kamikado:2013sia, Strodthoff:2016pxx}.

Another possible route towards real-time quantities is not to employ the above approaches directly in Minkowski space, but rather, to develop methods that allow for the analytic continuation of Euclidean data. A standard method used to extract the spectral function from Euclidean data emploies Bayesian inference to invert the integral relation between Euclidean correlation functions and spectral functions. In order to overcome some difficulties of the Maximal Entropy Method (MEM) \cite{Jarrell:1996rrw}, a novel Bayesian method was developed recently in Ref.~\cite{Burnier:2013nla} and applied for example to extract spectral properties red from bottomonium correlators in Ref.~\cite{Kim:2014iga} and solutions of the quark Dyson-Schwinger equation obtained in Landau gauge \cite{Fischer:2017kbq}. In the context of the $\Phi$-derivable expansion scheme (also referred to as the two-particle irreducible (2PI) formalism), knowledge of the external propagator, which fully displays the (linear) symmetries of the theory, requires  the resolution of a Bethe-Salpeter equation. Since the latter equation is more easily solved in Euclidean space (see for instance \cite{Carrington:2013jta}), analytic continuation techniques are of great help in this case. 

In this paper, we would like to investigate the analytic continuation method of Ref.~\cite{VS_Pade} which constructs multipoint Pad{\'e} approximants in the form of finite continued fractions. We test this method using various benchmarks. First, we consider the two-loop $\Phi$-derivable approximation for a one-component scalar field at zero temperature for which Euclidean data are already available \cite{Fejos:2011zq} and Minkowski data are relatively easy to generate using dispersion techniques. Using this benchmark, we test to what extent the pole mass and more generally the spectral function determined directly in Minkowski space can be reproduced from the Euclidean solution. Our test shows also that we have some control on the choice of the appropriate analytic continuation (among the infinitely many possible choices). As a second test, we use  simple one-loop perturbative formulas and investigate the ability of the continuation method to extract (physical) information hidden in the second Riemann sheet.   

We then apply the Pad{\'e} analytic continuation method on the previously obtained Euclidean space solution of Refs.~\cite{Marko:2013lxa, Marko:2015gpa} to assess how much pole masses differ from zero-momentum masses. A similar question was recently addressed in Ref.~\cite{Helmboldt:2014iya} in the context of the quark-meson model using the FRG approach. In our case, such a comparison is motivated by the fact that, in our previous Euclidean studies of the $O(4)$ model, the parametrization was done at $T=0$ based, for simplicity, on the pion and sigma zero-momentum masses of the external propagator which are directly accessible from the effective potential in the imaginary-time formalism. Since a more correct parametrization should involve the corresponding pole masses, we would like to quantify how much the parametrization was deformed by the use of the zero-momentum masses. We mention also that, recently, the Pad{\'e} analytic continuation method was applied in Ref.~\cite{Pilaftsis:2013xna} to compute the pole mass of the internal  propagator of the symmetry improved 2PI (SI2PI) formalism. In Ref.~\cite{Marko:2016wtw} we argued that, at least at two-loop order, the SI2PI formalism contains an inherent untamed infrared sensitivity in the broken phase, according to which the Euclidean propagator is not defined above some value of the volume of the system.\footnote{This infrared sensitivity can be avoided for some specific (sharp) UV regulators. However, the specificity of these UV regulators makes this removal of the IR sensitivity artificial in a sense, see Ref.~\cite{Marko:2016wtw} for more details.} Here we use the Pad{\'e} analytic continuation method to investigate to which extent this IR sensitivity affects the pole mass of the corresponding Minkowski propagator.

The organization of the paper is as follows. In Sec.~II, using dispersion techniques, we generate Minkowski data within the two-loop $\Phi$-derivable approximation for a $\varphi^4$ model at zero temperature, with or without spontaneous symmetry breaking. In Sec.~III, we present the multipoint Pad{\'e} continuation method and perform various tests using both the Minkowski space solution of Sec.~II and simple perturbative formulas. The continuation method is then used in Sec.~IV to assess the difference between pole and zero-momentum masses in the context of the $O(4)$ model, as well as the IR sensitivity of the pole mass in the two-loop SI2PI approximation. The conclusions of our study are presented in Sec.~V and some technicalities can be found in the Appendices.

\section{Minkowskian two-loop $\Phi$-derivable approximation in the $\varphi^4$ model\label{sec:2PIMink}}

In this section, we consider a relatively simple situation where the $\Phi$-derivable equations can be solved directly and accurately in Minkowski space. The obtained results shall then be used as a benchmark for the Pad{\'e} analytic continuation method to be presented in the next section. In what follows $Q=(q_0,\vec{q})$ denotes a four-momentum in Minkowski space and we introduce the notation
\[
\int_Q\equiv \int_{-\infty}^\infty\frac{d q_0}{2\pi}\int\frac{d q^3}{(2\pi)^3}\,.
\]

\subsection{Generalities}

The 2PI formalism gives typically access to a self-consistent equation for the two-point function of a given model, from which various observables can be determined, in principle. In contrast to the infinite tower of Dyson-Schwinger equations, the equation for the two-point function in the 2PI formalism is not coupled to higher $n$-point functions. The prize to pay is, however, that the equation contains infinitely many terms coming from the loop expansion of the 2PI effective action and then it needs to be truncated for any practical purpose. In the case of the $\varphi^4$ model, the two-loop truncation of the 2PI effective action leads to the following gap equation for the momentum dependent mass function $\bar M(Q)$ of the self-consistent propagator $\bar G(Q)\equiv i/(Q^2-\bar M^2(Q)+i\epsilon)$ in the presence of an arbitrary field expectation value $\phi$:
\be
\label{eq:gap}
\bar M^2(Q) = m_0^2+\frac{\lambda_2}{2}\phi^2+\frac{\lambda_0}{2}{\cal T}[\bar G]+\frac{\lambda^2}{2}\phi^2{\cal I}[\bar G](Q),
\ee
where the tadpole and bubble integrals are defined respectively as
\bea
\label{Eq:tad}
{\cal T}[G] & \equiv & \int_Q G(Q)\,,\\
{\cal I}[G_1,G_2](K) & \equiv & -i \int_Q G_1(Q) G_2(Q+K)\,,
\label{Eq:bubble}
\eea
with the shorthand notation ${\cal I}[G](K)\equiv {\cal I}[G,G](K)$. The physical value of the field $\phi$, denoted $\bar\phi$, is given in terms of the field equation. In the two-loop order truncation, it reads
\be
0 = \bar\phi\left(m_2^2 + \frac{\lambda_4}{6}\bar\phi^2 + \frac{\lambda_2}{2}{\cal T}[\bar G] + \frac{\lambda^2}{6}{\cal S}[\bar G] \right),
\label{eq:field}
\ee
where the setting-sun integral at vanishing external four-momentum is defined as
\bea
{\cal S}[G_1,G_2,G_3] & \,\,\equiv \,\, & -i\int_Q \int_K G_1(Q) G_2(K) G_3(Q+K)\,,
\label{Eq:SS}
\eea
with a similar shorthand notation as above, ${\cal S}[G]\equiv {\cal S}[G,G,G]$. Let us mention that the field equation may have multiple solutions. The solution $\bar\phi$ we are interested in is the one that minimizes the effective potential. The latter can also be computed within the 2PI formalism, but we shall not recall its expression here.

As discussed in Ref.~\cite{Berges:2005hc}, a total of five counterterms are needed in order to absorb the divergences present in Eqs.~(\ref{eq:gap}) and (\ref{eq:field}). These counterterms can be expressed in terms of two renormalized quantities, a mass parameter $m^2$ and a coupling $\lambda,$ as well as a renormalization scale. Expanding the propagator around an auxiliary propagator $G_0(Q)=i/(Q^2-M_0^2+i\varepsilon),$ in which $M_0$ plays the role of the renormalization scale, the counterterms were explicitly determined in Ref.~\cite{Patkos:2008ik} and employed numerically in Ref.~\cite{Fejos:2011zq}. Similar counterterms, although determined at some fixed temperature in the symmetric phase, have been used in \cite{Arrizabalaga:2007zz,Marko:2012wc, Marko:2013lxa, Marko:2015gpa}. In the present study, we use the counterterms of Ref.~\cite{Patkos:2008ik} and, in Appendix~\ref{sec:renorm}, we show the formal relation between the renormalization method of Ref.~\cite{Patkos:2008ik} and that of Ref.~\cite{Berges:2005hc}, which is formulated in terms of various two-point and four-point functions. The explicitly finite gap and field equations obtained after renormalization are
\begin{subequations}
\label{Eq:gap-field_F}  
\bea
\bar M^2(Q)&=&m^2+\frac{\lambda}{2}\left(\phi^2+{\cal T}_{\rm F}[\bar G]\right)+\frac{\lambda^2}{2}\phi^2 {\cal I}_{\rm F}[\bar G](Q)\,,\label{Eq:gap_F}\\
0 &=& \bar\phi \left[m^2+\frac{\lambda}{6}\bar\phi^2+\frac{\lambda}{2}{\cal T}_{\rm F}[\bar G]+\frac{\lambda^2}{6} {\cal S}_{\rm F}[\bar G]\right]. \label{Eq:field_F}
\eea
\end{subequations}
The finite parts of the integrals, denoted by the subscript `F', read
\begin{subequations}
\label{Eq:TIS_F}	
\bea
{\cal T}_{\rm F}[G] & \equiv & \int_Q G_{\rm r}(Q)\,,\label{Eq:tad_F}\\
{\cal I}_{\rm F}[G](Q) & \equiv & {\cal I}[G](Q) - {\cal I}[G_0]\,, \label{Eq:bubble_F}\\
{\cal S}_{\rm F}[G] & \equiv & {\cal S}[\delta G] + 3 {\cal S}[\delta G,\delta G, G_0] + 3 \int_Q G_{\rm r}(Q) {\cal I}_{\rm F}[G_0](Q)\,, \label{Eq:SS_F} 
\eea
\end{subequations}
where ${\cal I}[G_0]\equiv {\cal I}[G_0](Q=0)$ and, as a result of expanding $G$ around $G_0$, one has
\begin{subequations}
\label{Eq:dG_Gr_Ml}
\bea
\delta G(Q) &=& G(Q) - G_0(Q)\,,\\
G_{\rm r}(Q) &=& \delta G(Q) - i\left(M_0^2-M_{\rm l}^2-\frac{\lambda^2}{2}\phi^2 {\cal I}_{\rm F}[G_0](Q)\right)G_0^2(Q)\,,
\eea
\end{subequations}
with $M^2_{\rm l} \equiv m^2+\frac{\lambda^2}{2}\left(\phi^2+{\cal T}_{\rm F}[G]\right)$. Each of the three terms in the r.h.s. of Eq.~\eqref{Eq:SS_F} is UV finite.

\subsection{Use of dispersion relations \label{ss:disp_rel}}

To put the system (\ref{Eq:gap-field_F}) in a convenient form in view of its numerical resolution, we now make use of dispersion relations. We first focus on the momentum dependent bubble integral ${\cal I}_{\rm F}[G](K)$. A standard method for dealing with such an integral in Minkowski space is to use the spectral representation for the propagator and a subtracted dispersion relation (see {\it e.g.} Ref.~\cite{Bjorken_Drell_RQF}) relating the real and imaginary parts of the integral. The spectral representation for the propagator is
\be
G(Q) = \int_0^\infty \frac{d s}{2\pi}\frac{i\rho(s)}{Q^2-s+i\varepsilon}\,,
\label{eq:spectral-rep}
\ee
with the spectral function defined as the (real) function $\rho(Q^2)=2\Im[i G(Q)].$ In the two-loop approximation considered here, the leading contribution to the propagator at large $Q$ is entirely given by the tree-level term, $G(Q)\sim i/Q^2$. It then follows from Eq.~(\ref{eq:spectral-rep}) that the spectral function obeys the following sum rule\footnote{This relation still holds beyond the present approximation for the spectral function associated with the bare propagator, in the presence of an ultraviolet regulator. It needs to be modified by a renormalization factor for the spectral function associated to the renormalized propagator.}
\be
\int_0^\infty\frac{d s}{2\pi} \rho(s) = 1\,.
\label{Eq:sumrule}
\ee
We note also that $\rho(s)\sim 1/s^2$ at large $s$ in the approximation at hand (if $\bar\phi\neq 0$).

Using Eq.~\eqref{eq:spectral-rep} in a particular integral containing the self-consistent propagator is helpful because the integral over the momentum becomes perturbative and can be carried out using standard techniques, yielding an integral kernel in the rewritten expression of the original integral. In particular, we can use this rewriting in order to evaluate the imaginary part of the bubble integral ${\cal I}_{\rm F}[G](Q)$. Using Eq.~\eqref{eq:spectral-rep} in the integral for ${\cal I}[G](Q)$, carrying out the momentum integration, and taking the imaginary part, while noting that $\Im {\cal I}_{\rm F}[G](Q)=\Im {\cal I}[G](Q)$, one obtains 
\beq
\label{eq:ImG}
\Im {\cal I}_{\rm F}[G](Q)=\int_0^\infty\frac{d s_1}{2\pi}\int_0^\infty\frac{d s_2}{2\pi}\rho(s_1)\rho(s_2)\Im {\cal I}_0[G_1,G_2](Q)\,,
\eeq
where $G_j(Q)=i/(Q^2-s_j+i\varepsilon),$ $j=1,2$. The UV finite kernel $\Im {\cal I}_0[G_1,G_2](q)$ is the imaginary part of the perturbative bubble with mass squares $s_1$ and $s_2.$ It is given in terms of the K{\"a}ll{\'e}n function $\lambda(x,y,z)=(x-y-z)^2-4yz$ as
\be
\Im {\cal I}_0[G_1,G_2](Q)=-\frac{\sqrt{\lambda(Q^2,s_1,s_2)}}{16\pi Q^2}\Theta\big(Q^2-(\sqrt{s_1}+\sqrt{s_2})^2\big),
\ee
with $\Theta(x)$ the Heaviside step function. 

We could obtain the real part of ${\cal I}_{\rm F}[G](Q)$ in the same fashion. It is however more efficient to make use of the dispersion relation that connects it to its imaginary part. Since the real part of the finite bubble integral grows logarithmically in the approximation at hand, we need a once-subtracted dispersion relation (see Appendix~\ref{app:disp} for more details):
\beq
\label{eq:dispRel_def}
\Re {\cal I}_{\rm F}[G]\left(\sqrt{s}\right) = {\cal I}_{\rm F}[G]+\frac{s}{\pi}{\cal P}\int_0^\infty d s'\frac{\Im {\cal I}[G]\big(\sqrt{s'}\big)}{s'(s'-s)}\,,
\eeq
where we have again used that $\Im {\cal I}_{\rm F}[G](Q)=\Im {\cal I}[G](Q)$ as well as $\Re {\cal I}_{\rm F}[G]={\cal I}_{\rm F}[G]$, with our notational convention ${\cal I}[G]={\cal I}[G](Q\equiv 0)$. Each term in the previous relation is finite. The subtracted piece ${\cal I}_{\rm F}[G]$ can be conveniently computed using Eq.~\eqref{eq:spectral-rep} for $G$. We find
\beq
{\cal I}_{\rm F}[G] & = & \int_0^\infty\frac{d s_1}{2\pi}\int_0^\infty\frac{d s_2}{2\pi}\rho(s_1)\rho(s_2) {\cal I}_0[G_1,G_2]-{\cal I}[G_0]\nonumber\\
& = & \int_0^\infty\frac{d s_1}{2\pi}\int_0^\infty\frac{d s_2}{2\pi}\rho(s_1)\rho(s_2) {\cal I}_{0,\rm F}[G_1,G_2]\,,
\label{Eq:local_IF}
\eeq
where $ {\cal I}_{0,\rm F}[G_1,G_2]= {\cal I}_0[G_1,G_2]-{\cal I}[G_0]$ and we have used the sum rule (\ref{Eq:sumrule}) to bring ${\cal I}[G_0]$ under the integral. The difference $ {\cal I}_{0,\rm F}[G_1,G_2]$ of pertubative bubble integrals is UV finite. It can then be computed in any regularization scheme, with the result
\be\label{Eq:ReIF_zm}
{\cal I}_{0,\rm F}[G_1,G_2]=\frac{(s_1-s_2)^{-1}}{16\pi^2}\left(s_1\ln\frac{s_1}{e M_0^2} - s_2\ln\frac{s_2}{e M_0^2}\right).
\ee

We now turn our attention to the principal value integral appearing in Eq.~\eqref{eq:dispRel_def}. Since this integral depends on which interval the imaginary part of the bubble has a support, we have to say a few words on the form of the spectral function. In the present section we restrict ourselves to spectral functions which have a singular part, corresponding to a real pole of the propagator. In this case one can write
\beq
\rho(s) = 2\pi Z\delta(s-\bar M^2_{\rm p})+\sigma(s)\,,
\label{eq:rhoPolesep}
\eeq
where the continuum part $\sigma(s)$, starting at the two-particle threshold, is given in Eq.~\eqref{eq:sigma} and the pole mass $\bar M_{\rm p}$ is defined as $\bar G^{-1}(Q=\bar M_{\rm p})=0$, or, equivalently, in terms of the gap mass $\bar M^2(Q)$, as
\beq
\label{eq:poleEq}
\bar M^2_{\rm p}=\bar M^2(\bar M^2_{\rm p})\,.
\eeq
Using that the imaginary part of the bubble integral, along with $\sigma,$ is non-zero only for $s>s_{\rm th}=4\bar M^2_{\rm p}$, one can write the real part of the finite bubble integral in the form
\be
\Re {\cal I}_{\rm F}[G]\left(\sqrt{s}\right) = {\cal I}_{\rm F}[G]+\frac{s}{\pi}{\cal P}\int_{s_{\rm th}}^\infty d s'\frac{\Im {\cal I}[G]\left(\sqrt{s'}\right)}{s'(s'-s)}\,,
\label{Eq:dispRel}
\ee
with ${\cal I}_{\rm F}[G]$ given in \eqref{Eq:local_IF}.  

Finally we discuss the evaluation of the finite tadpole and setting-sun integrals  ${\cal T}_{\rm F}[G]$ and ${\cal S}_{\rm F}[G]$ defined in \eqref{Eq:tad_F} and \eqref{Eq:SS_F}. The simplest way to compute them is by using the Euclidean propagator obtained through analytic continuation from \eqref{eq:spectral-rep} as
\beq
G_{\rm E}(Q_{\rm E})=\int_0^\infty\frac{d s}{2\pi}\frac{\rho(s)}{Q^2_{\rm E}+s}\,.
\label{Eq:spectral-rep_E}
\eeq
For instance, the unsubtracted tadpole ${\cal T}[G]$ that enters ${\cal T}_{\rm F}[G]$ can be written, owing to a Wick rotation, as ${\cal T}[G]=\int_{Q_{\rm E}} G_{\rm E}(Q_{\rm E})\equiv {\cal T}_{\rm E}[G_{\rm E}]$. Similarly, one finds $\smash{{\cal S}[G]=-\int_{Q_{\rm E}}\int_{K_{\rm E}} G_{\rm E}(Q)G_{\rm E}(K)G_{\rm E}(Q+K)\equiv - {\cal S}_{\rm E}[G_{\rm E}]}$. With the Euclidean version of \eqref{Eq:dG_Gr_Ml}, ${\cal T}_{\rm F}[G]$ and ${\cal S}_{\rm F}[G]$ are then evaluated with an appropriate numerical cutoff.

\subsection{Results}
The iterative numerical algorithm for the resolution of the system (\ref{Eq:gap-field_F}) is described in detail in Appendix~\ref{app:algo}. Here we concentrate on describing the results obtained in the broken symmetry phase of the model. We explore in Sec.~\ref{ss:param} the region of the parameter space ($m^2,\lambda$) where such type of solution exists. All dimensionfull quantities are measured in units of $M_0$ throughout this paper, that is in the numerical code $M_0=1.$ 

\subsubsection{Minkowskian solution}

\begin{figure}
\begin{center}
\includegraphics[width=0.485\textwidth]{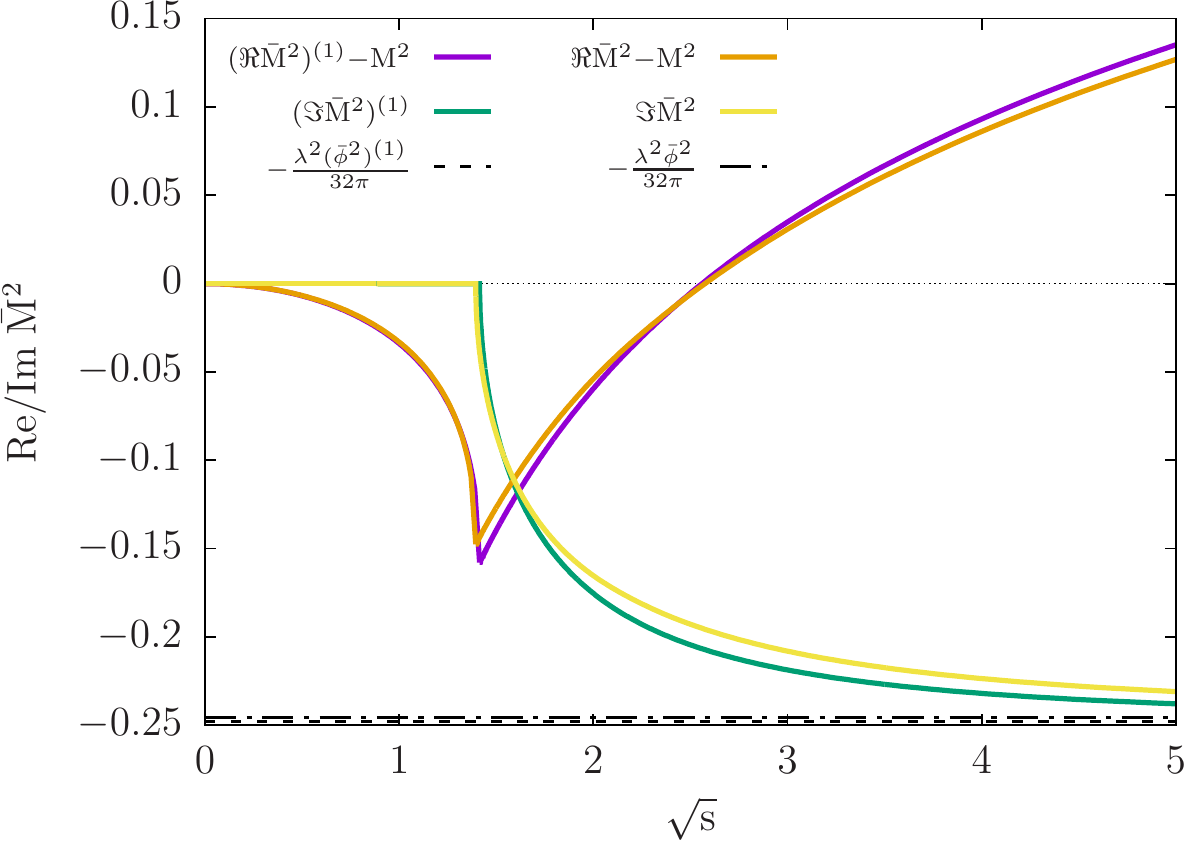}\hspace{0.5cm}\includegraphics[width=0.485\textwidth]{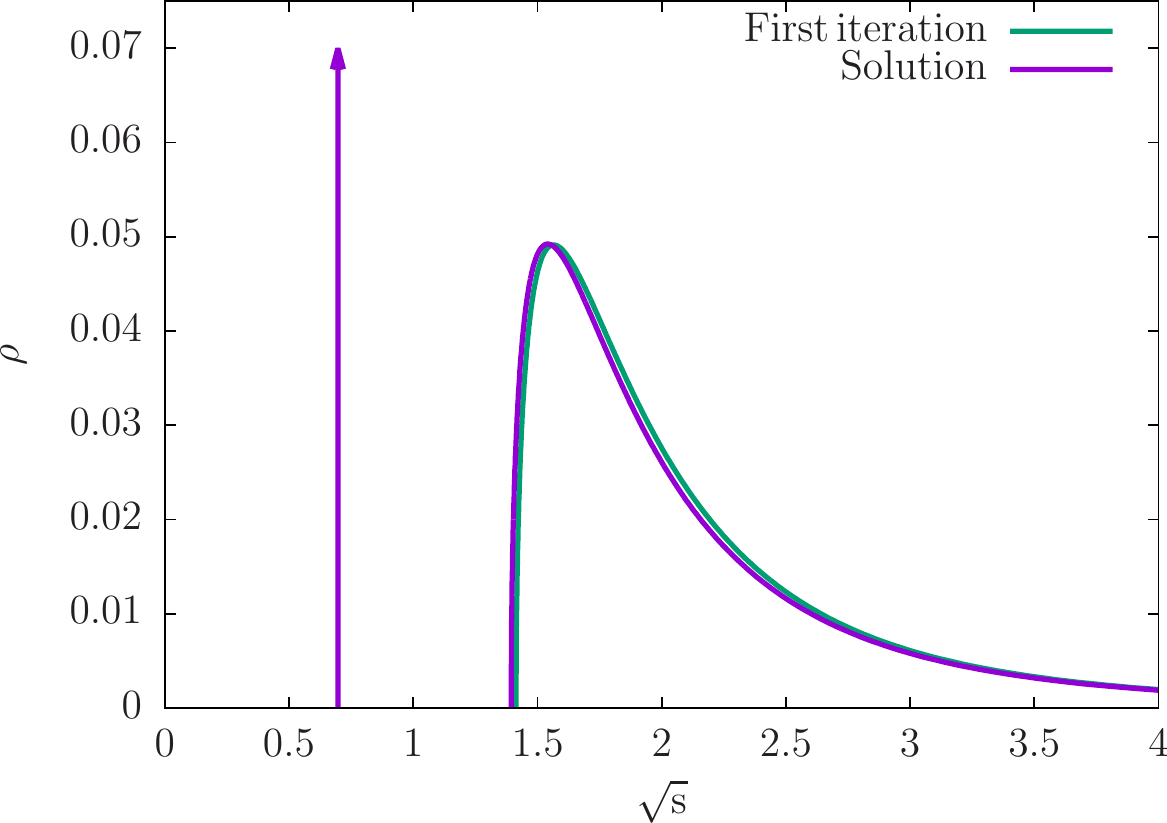}
\caption{Left panel: real and imaginary parts of the self-energy for a broken phase solution with $\lambda=15$ and $m^2=-0.28666$ obtained by solving \eqref{Eq:gap-field_F} and characterized by $\bar M^2(Q=0)\equiv M^2=0.5,$ $\bar M^2_{\rm p}=0.4862,\,\bar\phi=0.3312$ and $Z=0.96904$, compared with a perturbative expression evaluated with a mass squared equal to $\bar M^2(Q=0)$ and a field value $\bar\phi^{(1)}=-0.3328$, obtained in the first iteration (in the plot the value of the field is illustrated by the value of the limit $\Im \bar M^2(Q\to\infty)$). Right panel: The spectral functions corresponding to the first iteration and to the solution of the 2PI equations. \label{Fig:SE-rho_res}}
\end{center}
\end{figure}

A typical broken phase solution of the gap and field equations \eqref{Eq:gap-field_F} can be seen in Fig.~\ref{Fig:SE-rho_res}: the left panel shows the real and imaginary part of the self-energy, while the right panel shows the pole and the continuum part of the corresponding spectral function. We compare these quantities to perturbative ones obtained using a tree-level spectral function $\rho(Q)=2\pi\delta(Q^2-M^2)$ to evaluate the integrals, with $M^2\equiv\bar M^2(Q=0)$. We see that even for relatively large value of the coupling the 2PI self-energy is not much different from a perturbative one. This is due to the fact that in the two-loop approximation the momentum dependence of the self-energy is logarithmic and that the mass $M^2\equiv\bar M^2(Q=0))$ is then very close to $\bar M_{\rm p}^2.$ Note however that resummation of a perturbative series is needed in order to know the value of $M^2$, hence the comparison tells that the momentum dependence of the self-energy is similar to a perturbative one.\footnote{Another possibility is to parameterize the system directly in terms of $M^2$, with however the important subtlety that two physically distinct systems (one in the broken phase and one in the symmetric phase) can lead to the same parameters $(M^2,\lambda)$, see the discussion in Appendix~\ref{app:algo}.}

\subsubsection{Parametrization \label{ss:param}}

In previous studies \cite{Marko:2012wc,Reinosa:2011ut}, we analyzed the parameter space of the model at $\smash{T=0}$ and how it was divided into regions corresponding to systems that displayed a symmetric phase and systems that displayed a broken phase. In those studies, the renormalization was carried out in the symmetric phase at a large enough fixed temperature $T_\star$ that played the role of the renormalization scale. It is interesting to see how this discussion appears in the present scheme where the renormalization is performed directly at $T=0$, where the parameters $m^2$ and $\lambda$ do not have the same meaning as in previous studies, and where the role of the renormalization scale is played by $M_0$ rather than $T_\star$.

Apart from the occurrence of the above mentioned two regions of the parameter space, the detailed investigation of Ref.~\cite{Marko:2015gpa} shows that, within a given 2PI approximation, one could have physical and unphysical branches of solutions to the gap equation (as $\phi$ is varied), which could merge for some values of the field. In this case, there is a region of $\phi$ where the gap equation admits no solution. This could result in the absence of solution to the coupled system of gap and field equations if the would-be $\bar\phi$ were to be engulfed by the region of $\phi$ over which the gap equation has no solution. We have shown in Ref.~\cite{Marko:2015gpa} that a localized approximation to the momentum dependent gap mass is useful to investigate the loss of solution. We now show that an unphysical solution to the full 2PI equations can indeed be found. However, for the sake of simplicity, we use the localized approximation introduced in Ref.~\cite{Marko:2015gpa} to find the region of parameters where a loss of solution happens.\\

\begin{figure}
\begin{center}
\includegraphics[width=0.48\textwidth]{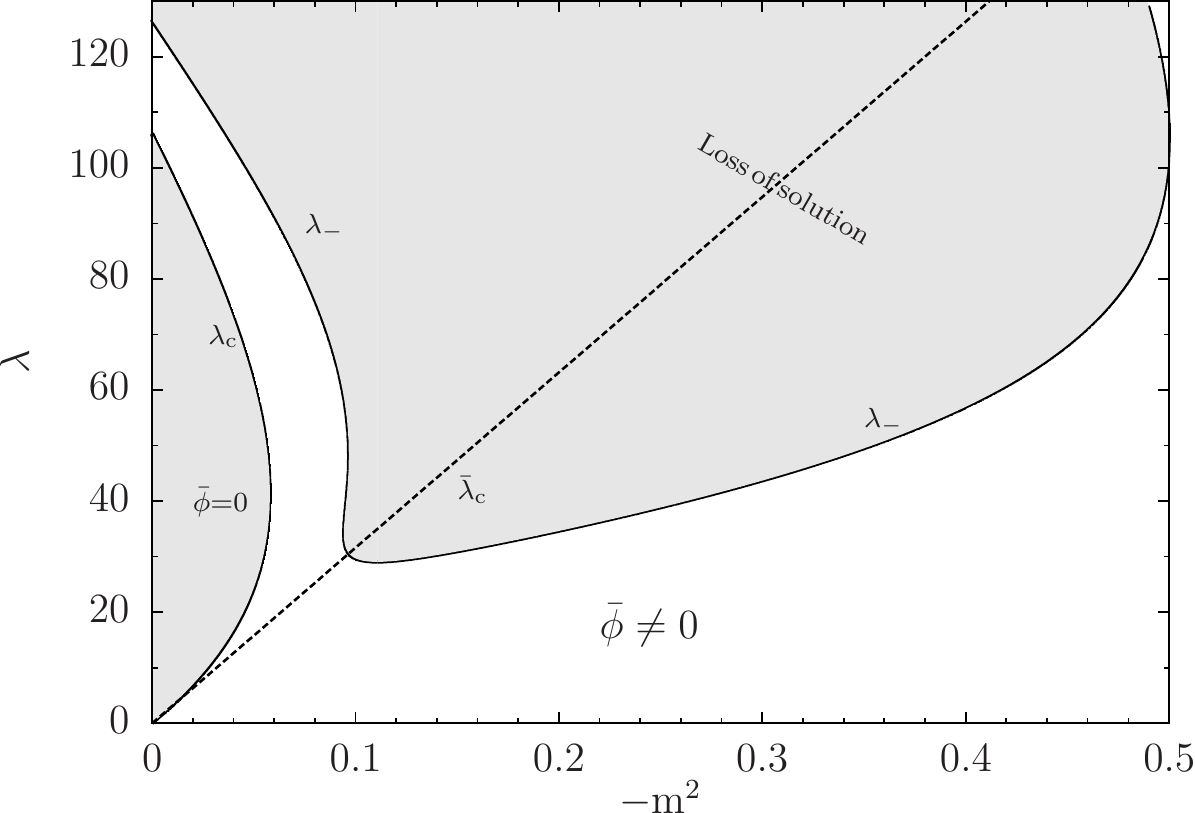}\includegraphics[width=0.5\textwidth]{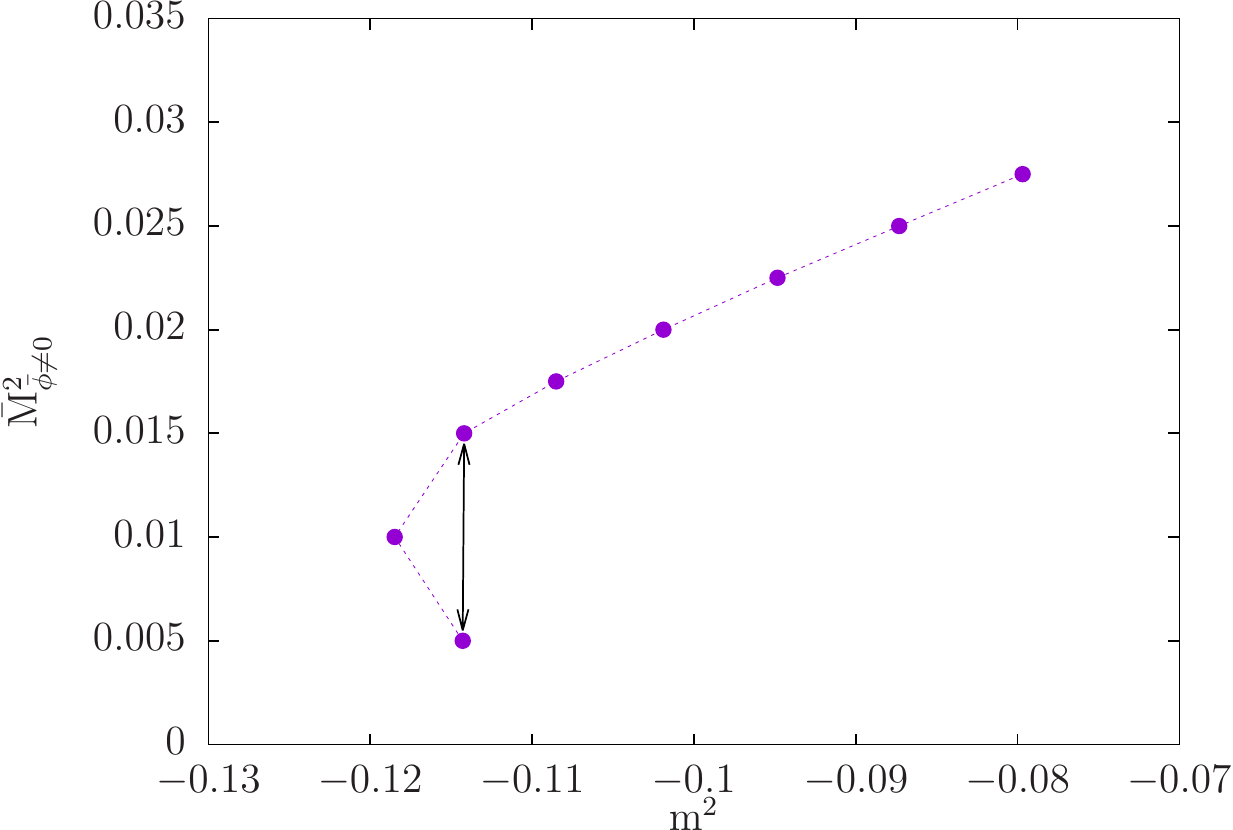}
\caption{Left panel: The parameter space in the localized approximation. The $\lambda_{\rm c}$ and $\bar \lambda_{\rm c}$ curves coincide with the ones of the full 2PI case, while $\lambda_{-}$ represents an approximation to the boundary of the region where a loss of solution is observed in the full 2PI case. Right panel: The gap mass of the broken phase solution $\bar M_{\bar\phi\neq0}^2$ as a function of the renormalized mass parameter $m^2.$ The results are obtained at a fixed $\lambda=34$ using the $M^2$ parametrization discussed in Appendix~\ref{app:algo}. The fact that the curve is multivalued in some $m^2$ range signals the appearance of an unphysical solution (lower branch) there. At this value of the coupling the loss of solution occurs in the region $m^2<-0.12$.\label{Fig:nonph}}
\end{center}
\end{figure}

Our analysis reveals the existence of three curves which, as shown in the left panel of Fig.~\ref{Fig:nonph}, delimit different regions of the $m^2,\,\lambda$ parameter space. The physically most relevant one, denoted $\lambda_{\rm c}(m^2)$, separates the region where only a symmetric phase exists from the region where there is potentially a broken phase solution. This curve is a critical line in parameter space, in the sense that, for these parameters, the physical solution is at $\bar\phi=0$ with no $\bar\phi\neq0$ solution and the curvature of the potential at $\phi=0$ is vanishing. The condition of vanishing curvature thus defines the $\lambda_{\rm c}(m^2)$ curve through the relation
\beq
m^2+\frac{\lambda_{\rm c}}{2}{\cal T}_{\rm F}[\bar G_{\phi=0}]+\frac{\lambda_{\rm c}^2}{6}{\cal S}_{\rm F}[\bar G_{\phi=0}]=0\,.
\eeq
The second line, denoted $\lambda_{-}(m^2)$, delimits the region where a loss of solution occurs from the region where the coupled gap and field equations do admit a solution. We expect such a region based on our previous investigations in Ref.~\cite{Marko:2015gpa}. As an example, the left panel of Fig.~\ref{Fig:nonph} shows a value of the coupling ($\lambda=34$) at which the solution is lost for $m^2<-0.12,$ while for $m^2>-0.12$, $\bar M_{\bar\phi\neq0}^2(m^2)$ is multivalued. Multivaluedness is usually the signal of the appearance of unphysical branches, which can collide with the physical branch and lead to a loss of solution. In our case, the physical branch is always the upper one. This branch continuously connects to the $\phi=0$ solution (in case of parameters where the solution exists for any $\phi\geq0$), where the unphysical branch disappears and only the physical solution remains.\footnote{Actually at all $\phi$, including the vanishing field limit, there exists yet another unphysical branch, where the gap mass is exponentially large. We disregard this solution of the gap equation when discussing physical and unphysical branches.} It is important to note here that, in order to find the unphysical branch of solution, it was very helpful to reparametrize our equations in terms of the local part of the self-energy $\bar M^2(Q=0)$, rather than $m^2$, as explained in Appendix~\ref{app:algo} around \eqref{Eq:M2}. For further details on unphysical branches of solutions and loss of solution we direct the reader to Ref.~\cite{Marko:2015gpa}. In principle, the $\lambda_{-}(m^2)$ curve can be obtained in the full 2PI case by searching, at each value of $m^2$, for the value of the coupling where a loss of solution to the coupled equations occurs, as shown in the right panel of Fig.~\ref{Fig:nonph}. Since this is a tedious procedure, as for large $\lambda$ the convergence of the iterations is slow and strongly initial condition dependent, we resort to the much simpler localized approximation used in Ref.~\cite{Bordag:2000tb,Reinosa:2011cs,Marko:2015gpa}, which qualitatively reproduces the solution of the full 2PI equations. In this approximation the full self-energy is replaced by its zero momentum value resulting in a tree-level type propagator with a self-consistent mass. Finally we define $\bar\lambda_{\rm c}(m^2)$ as the curve where the gap mass at $\phi=0$ vanishes. For $\lambda < \bar\lambda_{\rm c}$ the gap equation admits no solution at $\phi=0$ which also makes the potential unreachable there. Using the gap equation \eqref{Eq:gap_F} at $\phi=0$ and that ${\cal T}_{\rm F}[\bar G]\big|_{\bar M^2=0}=M_0^2/(16\pi^2)$ one can easily see that $\bar\lambda_{\rm c}(m^2)=-32 \pi^2 m^2/M_0^2\,.$

\section{Analytic continuation of Euclidean solutions using Pad{\'e} approximants}

In this section we would like to use Pad{\'e} approximants in order to analytically continue functions known only at a finite number of points in the complex plane. We employ the multipoint Pad{\'e} approximant (see Refs.~\cite{VS_Pade} and \cite{Baker}) calculated for a function known at $N$ complex points $z_i$, $f(z_i)=u_i,$ $i=1 \dots,N,$ using a finite continued fraction. Using the notation $\frac{1}{1+}x\equiv\frac{1}{1+x}$, the finite continued fraction is written in the form
\be
\label{Eq:Pade_CN}
C_N(z)=\frac{a_1}{1+}\,\frac{a_2(z-z_i)}{1+}\cdots\frac{a_N(z-z_{N-1})}{1}\,,
\ee
and the task is to determine its $N$ coefficients $a_i$ from the conditions $C_N(z_i)=u_i.$ An elegant and efficient way to achieve this is by recursion: the coefficient are obtained as $a_i=g_i(z_i)$, by defining\footnote{In the second relation it is understood that $z$ is part of the set of discrete points $z_i.$}
\begin{subequations}
\label{Eq:recursion}
\bea
g_1(z_i)&=&u_i,\quad i=1,\dots,N\,,\\
g_p(z)&=&\frac{g_{p-1}(z_{p-1})-g_{p-1}(z)}{(z-z_{p-1})g_{p-1}},\ \ p\geqslant 2\,.
\eea
\end{subequations}
Working out explicitly the condition $a_i=g_i(z_i)$ for a few values of $i$ one sees that basically one needs to construct a triangular matrix $t_{i,j}$ using the recursion $t_{i,j}=(t_{i-1,i-1}/t_{i-1,j}-1)/(z_j-z_{i-1}),$ for $j=2,\dots,N$ and $i=2,\dots,j$, starting from its first row $t_{1,j}=u_j,$ $j=1,\dots,N$. 

The finite continued fraction can be written as a rational function: $C_N(z)=A(z)/B(z)$, which we do not give here because we do not use that form. We only mention that the polynomials $A(z)$ and $B(z)$ are both of order $(N-1)/2$ for $N$ odd and of order $(N-2)/2$ and $N/2$, respectively, for $N$ even. This means that with increasing $N$ two Pad{\'e} sequences of the type $P_k^k(z)$ (for $N=2k+1, k\geqslant 0$) and $P_{k+1}^k(z)$ (for $N=2(k+1), k\geqslant 0$) are generated. When all the continued fraction coefficients are nonnegative it is known (see Ref.~\cite{Bender}) that $P_k^k(z)$ decreases, while $P_{k+1}^k(z)$ increases monotonically with $k$. The first sequence has a lower bound, while the second sequence has an upper bound which for $k\to \infty$ is a Stieltjes function $F(z)$, that is a function of the form $F(z)=\int_0^\infty d t\rho(t)/(1+z t),$ with $\rho(t)\geqslant 0$ in the domain of integration.\\

We shall use the above method to obtain the propagator $\bar G(Q)$ for $Q^2>0$, from the knowledge of the same propagator at a finite number of negative $Q^2=-Q_{\rm E}^2$ values in the Euclidean domain. In practice we fix $\vec{q}$ (to zero in our case), define the original Pad{\'e} approximant as a function of the Matsubara frequency $\omega_n$ and, to carry out the analytic continuation, we evaluate it at $\omega_n=-i \omega+\varepsilon$, where the $\varepsilon\to0$ limit can be safely taken. We should of course keep in mind that the analytic continuation of a finite number of data is not unique and leads to various solutions that differ by their asymptotic behaviors as $|z|\to\infty$.\footnote{Knowing some data $d_1,\dots,d_n$ for a finite set of points $z_1,\dots,z_n$, and given an analytic continuation $f(z)$ of these data, which is such that $f(z_i)=d_i$, we can construct another analytic continuation of the same data as, for instance, $g(z)=f(z)+\prod_{j=1}^n (\exp\{i 2\pi z/z_j\}-1)$.} The ability of the method to reproduce the expected propagator needs then to be tested and we will do so by using the explicit Minkowskian solution obtained in the previous section. Another subtle point is that, by definition, the Pad{\'e} approximant has no branch cut and thus there is only one Riemann sheet to be considered. It is then a question how the method can allow us to access physical information usually hidden in the second Riemann sheet of the propagator. A related issued it the fact that the Pad{\'e} approximant does not obey the Schwarz reflection property, whereas the analytic propagator (\ref{eq:analytic}) obeys this property (${\cal G}(s^*)^*=-{\cal G}(s)$ with our conventions) in the first Riemann sheet, as it is easily checked.

\subsection{Test I: ability to access the spectral function \label{ss:test1}}
To test the quality of the Pad{\'e} analytic continuation we use two test functions: the Euclidean 2PI self-energy extracted from Eq.~\eqref{Eq:spectral-rep_E} using the spectral function obtained as the solution of \eqref{Eq:gap-field_F} and the zero temperature finite perturbative Euclidean bubble with square mass $M^2=0.5$:
\beq
\displaystyle
{\cal B}_{0,{\rm F}}[G_M](Q_{\rm E})=\frac{1}{16\pi^2}\left(2-\log\frac{M^2}{M_0^2}+\sqrt{1+\frac{4M^2}{Q^2_{\rm E}}}\log\frac{\sqrt{1+\frac{4M^2}{Q^2_{\rm E}}}-1}{\sqrt{1+\frac{4M^{2}}{Q^2_{\rm E}}}+1}\right).
\eeq
In both cases we compare the real and imaginary parts of the analytically continued quantity to the one computed in the Minkowski space, which in the latter case is $\bar M^{2}(Q)$. For the comparison in the 2PI case we use the parameters $m^2=-0.2611$ and $\lambda=5$, for which one obtains $M^2\equiv \bar M^2(Q=0)=0.5$ and $\bar \phi=0.5566.$ These agree with the values obtained with the numerical code used in Ref.~\cite{Fejos:2011zq} to solve the Euclidean version of the model in the present renormalization scheme.

\begin{figure}
\begin{center}
\includegraphics[width=0.485\textwidth]{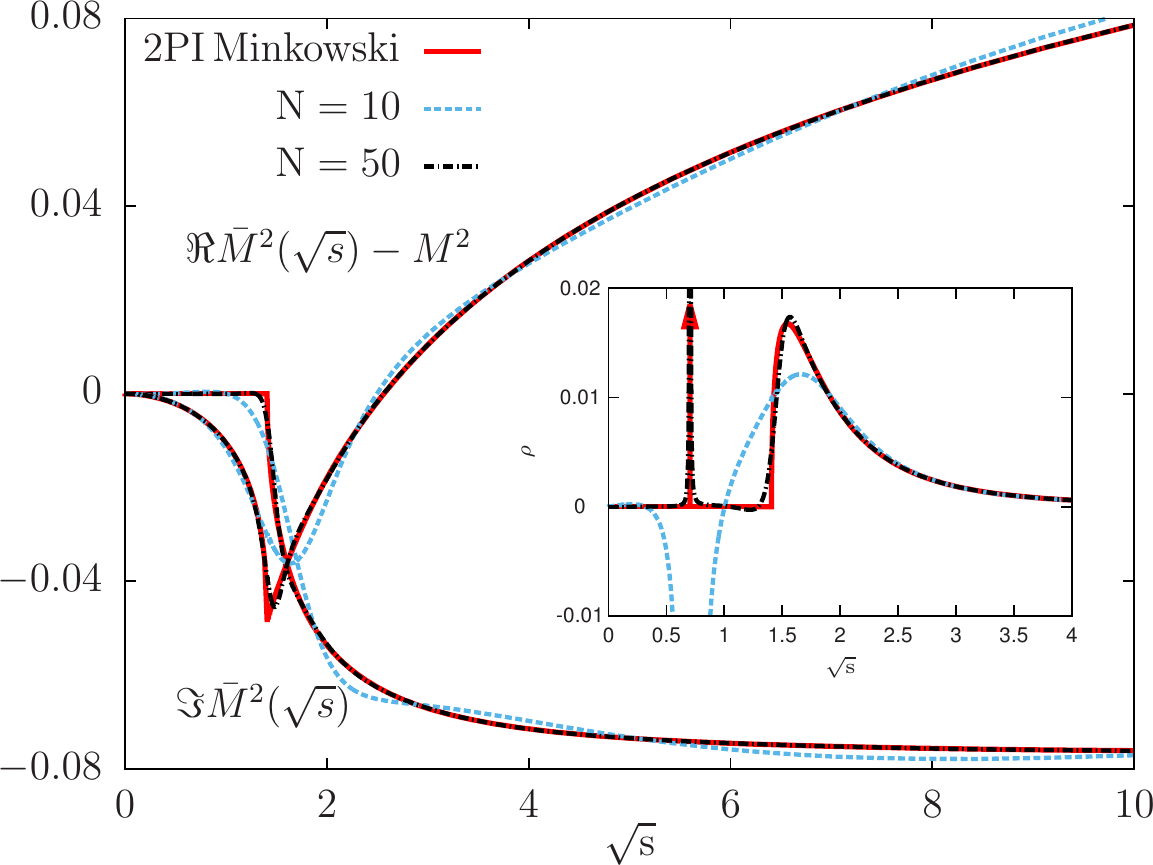}\caption{The real and the imaginary parts of the analytically continued Euclidean 2PI self-energy given by \eqref{Eq:spectral-rep_E} are compared to the corresponding quantities computed directly in Minkowski space from \eqref{Eq:gap-field_F}. The inset shows the same comparison for the spectral function. For Pad{\'e} analytic continuation we used $N=10$ and $N=50$ points in the $[0, 10M]$ interval of Euclidean momentum.\label{fig:PadeEg}}
\end{center}
\end{figure}

In Fig.~\ref{fig:PadeEg}, we show the Pad{\'e} analytically continued real and imaginary parts of the Euclidean 2PI self-energy and the corresponding spectral function compared to the quantities computed directly in Minkowski space. We see that while using $N=10$ points in \eqref{Eq:Pade_CN} leads to a noticeable error almost everywhere, except for the real part of the self-energy at small $\sqrt{s}$, the multipoint Pad{\'e} approximation using $N=50$ points has difficulties reproducing the Minkowski result only in a narrow neighborhood of the threshold. However, even with $N=10$ points, the analytical continuation gives a good approximation for the pole mass, $\bar M_{\rm p}=0.70387$, compared to the value $\bar M_{\rm p}=0.70392$ obtained from a direct calculation. Based on this observation, we can trust our results presented in Sec.~\ref{Sec:Comparison} concerning the comparison of zero-momentum and pole masses.

\begin{figure}
\begin{center}
  \includegraphics[width=0.485\textwidth]{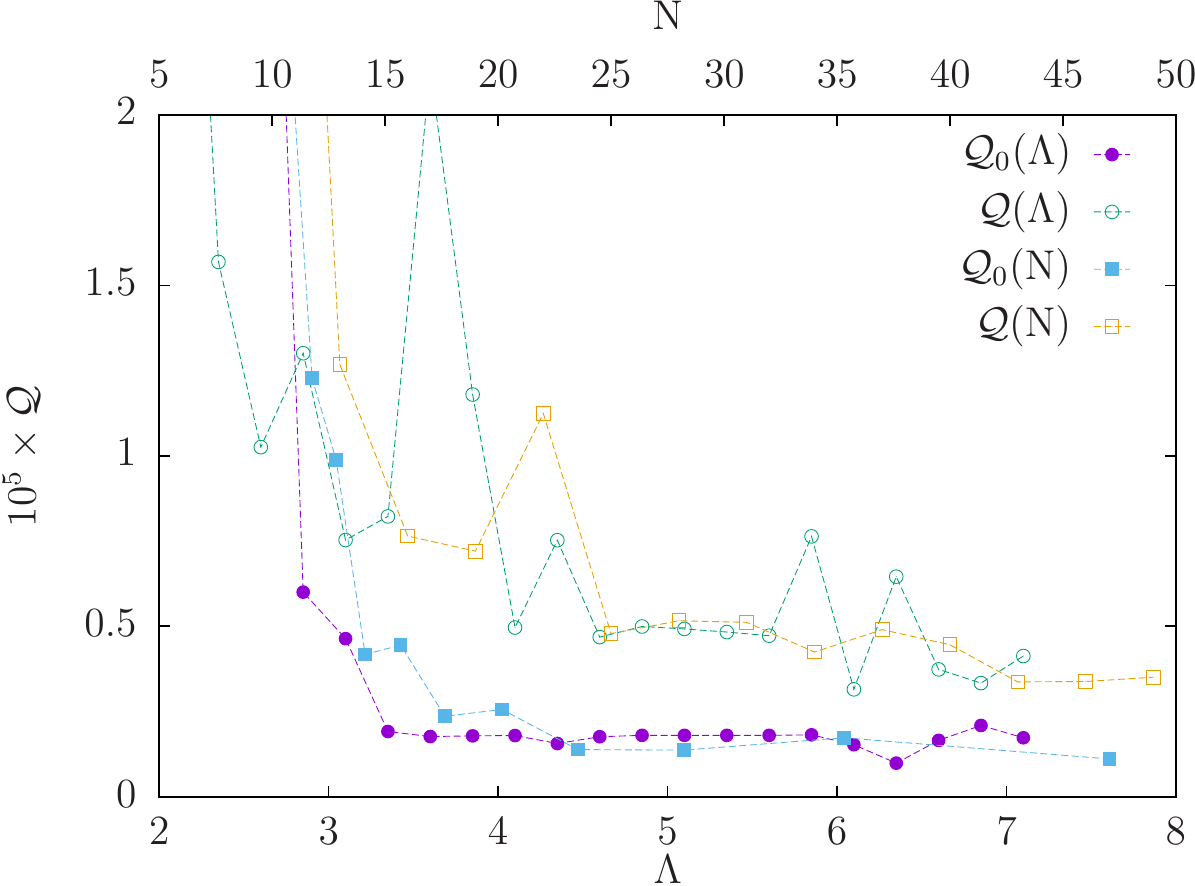}\caption{The quality of the Pad{\'e} analytic continuation of the Euclidean perturbative bubble (${\cal Q}_0$, filled symbols), and the Euclidean self-energy (${\cal Q}$, empty symbols). The circle symbols correspond to the bottom $x$-axis and were obtained using fixed number of sampling points ($N=50$) in an interval with variable upper limit $\Lambda$, while the square symbols correspond to the top $x$-axis and were obtained using $N$ sampling points in a fixed width interval with upper limit $\Lambda=10.$ \label{fig:PadeQ}}
\end{center}
\end{figure}

In order to quantify the quality of the analytic continuation over a significant region of momentum, we define the following Minkowski space integral
\beq
{\cal Q} = \int_{\sqrt{s}_{\rm min}}^{\sqrt{s}_{\rm max}}d q \big(F(q)-{\cal F}(q)\big)^2,
\eeq 
where $F$ is the original function and ${\cal F}$ its approximation obtained using analytic continuation, and with $\sqrt{s}_{\rm min/max}$ chosen appropriately. In Fig.~\ref{fig:PadeQ} we show this quantity ${\cal Q}$ as a function of the sampled region with fixed sampling frequency (bottom $x$-axis), and as a function of the number of sample points on fixed region (top $x$-axis). We see convergence both by increasing the sampled region and by increasing the number of points. While a perfect continuation would mean ${\cal Q}=0$ the curves do not tend to zero, which we attribute to the error arising from finite number representation and the inherent differences in the analytic properties of the approximated functions and the Pad{\'e} approximants. Also note that the combined results have some implications on the continuation of finite temperature data. The results corresponding to the bottom axis tell us that, at a certain temperature, there is a limiting Matsubara-frequency, over which including new frequencies does not improve the quality of the continuation. At the same time, the results associated to the top axis indicate that with fewer frequencies in the relevant range, {\it i.e.} by increasing the temperature, the quality of the continuation deteriorates.

\subsection{Test II: ability to access physical information in the second Riemann sheet}

We show now that the Pad{\'e} analytic continuation is capable of finding physically relevant complex poles in the case of an $O(4)$ symmetric model. Since we have 2PI results in Minkowski space only in the one-component case, we use as a benchmark the $T=0$ Minkowskian results of \cite{Chiku:1998kd} and \cite{Hidaka:2003mm} obtained within perturbation theory at one loop. We calculate the integrals of the sigma self-energy both in Euclidean and Minkowski spaces with the renormalization prescription of \cite{Chiku:1998kd} and then carry out the analytic continuation (cf. Ch.~6.3 of \cite{Brown}) of the Euclidean bubble integrals using Pad{\'e} approximants. We construct one Pad{\'e} approximant for each bubble integral, as it turns out that with this procedure we get better agreement with the Minkowskian results than in the case where a single Pad{\'e} approximant is associated to the entire Euclidean self-energy.

The physically relevant complex pole of the sigma propagator is on the 2nd Riemann sheet (see {\it e.g.} Sec.~1.3 of \cite{Badalyan}), which is accessed by crossing the real axis in between the thresholds of the pion and the sigma bubbles. In the present case this range is given in terms of the tree-level pion and sigma masses as $\sqrt{s}\in[2 m_{\pi,0}, 2 m_{\sigma,0}]$.

\begin{figure}
\begin{center}
\includegraphics[width=0.485\textwidth]{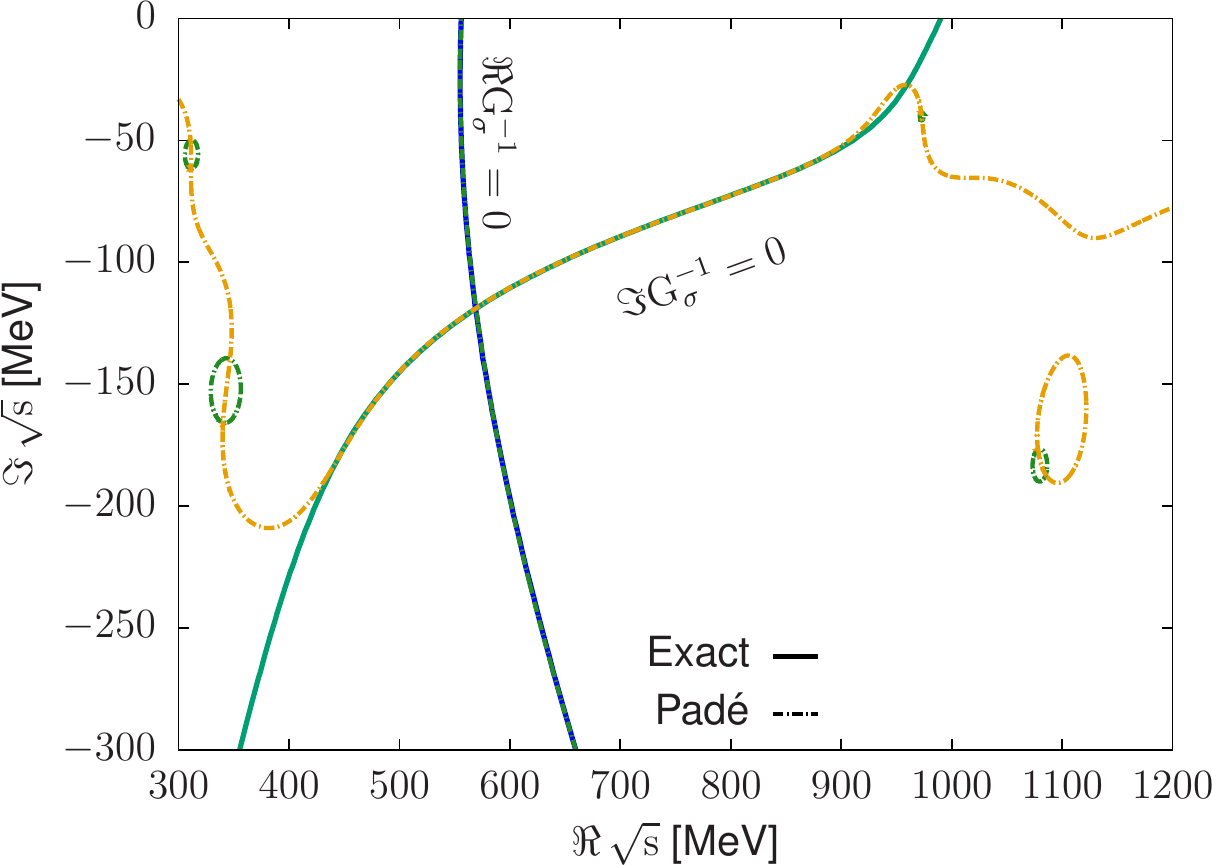}\caption{The zero contours of the real and imaginary parts of the inverse $\sigma$ propagator given by analytic expression on the 2nd Riemann sheet compared to the corresponding quantities obtained using multipoint Pad{\'e} approximation with $N=200$ points. The physical pole is well reproduced, although the Pad{\'e} approximation gives additionally eight spurious poles in the shown range.\label{fig:complexPole}}
\end{center}
\end{figure}

In search for the complex pole of the $\sigma$ propagator, we continue the Pad{\'e} approximant \eqref{Eq:Pade_CN} to complex values of its argument. With the parameters given in the first line of Table~I in Ref.~\cite{Chiku:1998kd}, we find a complex pole at $\sqrt{s}=569.25 - i\cdot 119.02$~MeV, using the exact formulas of the bubbles analytically continued to the 2nd Riemann sheet, and at $\sqrt{s}=569.32 - i\cdot 118.83$~MeV, using the multipoint Pad{\'e} approximation with $N=200$ points in the $[0, 3]$~GeV interval of Euclidean momentum. This finding is in line with the low temperature location of the pole II in the 2nd Riemann sheet shown in Fig.~6 of Ref.~\cite{Hidaka:2003mm}.\footnote{However, we could not reproduce the results shown in Fig.~2 of Ref.~\cite{Hidaka:2002xv}, where the scale of the imaginary axis seems to be off.} As in Fig.~3 of Ref.~\cite{Hidaka:2002xv}, the real part of this pole is close to the maximum of the spectral function, which is accurately reproduced with a Pad{\'e} approximant using $N=200$ points, in line with the findings of the previous subsection. However, a word of caution is in order, since as we show in Fig.~\ref{fig:complexPole}, far from the real axis the contours $\Re G_{\sigma}^{-1}(\sqrt{s})=0$ and $\Im G_{\sigma}^{-1}(\sqrt{s})=0$ are strongly distorted due to spurious poles of the Pad{\'e} approximants fitted to the bubble integrals. In the shown range there are eight fake poles, four to the left and four to the right of the physical one. While in our benchmark case we are lucky, as the position of the physical pole is well captured with the Pad{\'e} analytic continuation, one should always check whether the result obtained in the complex plane is influenced by non-physical poles of the Pad{\'e} approximant. 

The fact that the Pad{\'e} analytic continuation gives so accurately the coordinates of a complex pole on the 2nd Riemann sheet located far away from the real axis is a surprise to us in view of the fact that, as already mentioned, the Pad{\'e} approximant does not know about the existence of different Riemann sheets. When using Pad{\'e} approximants one can in principle mimic to some extent the procedure of the analytic continuation based on a known functional form. In order to do this, we continue the Euclidean bubbles to the real axis and then we construct new Pad{\'e} approximants for the real and imaginary part of the pion bubble for $\sqrt{s}>2m_{\pi,0}$ and a new Pad{\'e} approximant for the the sigma self-energy in the range $\sqrt{s}\in[2 m_{\pi,0}, 2 m_{\sigma,0}]$ where it is purely real. One can then continue these Pad{\'e} approximants to complex values. Actually, Fig.~\ref{fig:complexPole} was obtained in this way, and in the present case no significant change in the coordinates of the complex pole and of the zero contours was observed for $\Re \sqrt{s}<2m_{\sigma,0}$.

\section{Comparison of pole and zero-momentum masses in the O(N) model\label{Sec:Comparison}}

In our earlier works \cite{Marko:2013lxa, Marko:2015gpa}, we have studied the parametrization of the $O(4)$-model in regard to its application to light meson phenomenology. The parametrization was done using zero-momentum masses for simplicity. However, a correct parametrization requires, instead, the use of pole masses. Here we use the Pad{\'e} analytic continuation method in order to assess how different these masses can be. We should mention that this comparison only makes sense in a given renormalization scheme because the zero-momentum masses are scheme dependent whereas the pole masses are not.

Another subtlety comes from the fact that, in a given 2PI approximation, one can define two types of propagators, the so-called internal and external propagators. The internal propagator is the one considered in Sec.~\ref{sec:2PIMink} for the case of the two-loop approximation. However, one can define an external propagator, as we recall in the next subsection. Since the parametrization of Ref.~\cite{Marko:2013lxa, Marko:2015gpa} was done using the zero-momentum mass of the external propagator, in what follows we compare this mass with the pole mass of the same propagator. In doing so, we shall use a well motivated approximation for the external propagator which leads to the same expression in the two-loop and ${\cal O}(\lambda^2)$ truncations of the 2PI effective action, the two approximations used in Refs.~\cite{Marko:2013lxa, Marko:2015gpa}. For completeness, we shall also do this comparison in the case of the internal propagators which are different in the two types of truncations.

\subsection{External vs internal propagators}
In the 2PI formalism, the internal propagator $\bar G_\phi$ in the presence of a given field expectation value is obtained by solving the equation $0=\delta\Gamma[\phi,G]/\delta G_{G=\bar G_\phi}$, where $\Gamma[\phi,G]$ is the so-called 2PI effective action. Knowing $\bar G_\phi$, one can reconstruct the usual 1PI effective action as $\Gamma[\phi]=\Gamma[\phi,\bar G_\phi]$. From the latter, it is also possible to compute the propagator as ${\cal G}^{-1}=\delta^2\Gamma/\delta\phi^2$. Although these two definitions for the two-point function are equivalent in the absence of approximations, they differ in practice whenever a truncation of $\Gamma[\phi,G]$ is considered and the second definition is referred to as the external propagator. We refer to \cite{Berges:2005hc} for details on the relation between the two types of propagators and their respective renormalization. Here we recall the expression for the external propagator in the two-loop and ${\cal O}(\lambda^2)$ truncations, in the $N=1$ case. In Euclidean configuration space, one obtains
\beq
{\cal G}^{-1}_{\rm E}(x,y)=\left.\frac{\delta^2\Gamma[\phi,G]}{\delta\phi(x)\delta\phi(y)}\right|_{G=\bar G_\phi}-\frac{1}{2}\int_{z_1,z_2,z_3,z_4}\Lambda(x;z_1,z_2)\bar G_{\rm E}(z_1,z_3)\bar G_{\rm E}(z_4,z_2)V(z_3,z_4;y)\,,
\eeq
with
\beq
\left.\frac{\delta^2\Gamma[\phi,G]}{\delta\phi(x)\delta\phi(y)}\right|_{G=\bar G_\phi}={\cal G}_{\rm E,0}^{-1}(x-y)+\frac{1}{2}\left[\lambda_4\phi^2+\lambda_2\bar G_{\rm E}(0)\right]\delta(x-y)-\frac{\lambda^2}{6}\bar G_{\rm E}^3(x-y)\,,
\eeq
where we have restricted to homogeneous fields $\phi(x)=\phi$ and where $\Lambda(x;z_1,z_2)$ and $V(z_1,z_2;x)$ are proportional to $\phi$ and have tree-level contributions equal to $\lambda_2\phi\,\delta(x-z_1)\delta(x-z_2)$. In this work, we shall neglect all contributions to $\Lambda$ and $V$ beyond tree-level, which is the minimum needed if we want the two-loop external propagator to contain at least the same diagrams than the two-loop internal one. Higher order contributions, since they are proportional to $\phi^{2}$, lead only to logarithmic modifications of the momentum dependence of the self-energy. We thus expect those contributions not to modify by much the values of the pole and zero-momentum masses. This is an important simplification because, in general, obtaining the function $V$ requires to solve a Bethe-Salpeter equation, which we avoid here without affecting our conclusions in a dramatic way. Putting these pieces together and Fourier transforming, we arrive at ${\cal G}^{-1}_{\rm E}(Q_{\rm E})\equiv Q_{\rm E}^2+\hat{M}_{\rm E}^2(Q_{\rm E})$, with
\beq
\hat{M}_{\rm E}^2(Q_{\rm E})=\delta Z_2 Q_{\rm E}^2+m_2^2+\frac{\lambda_4}{2}\phi^2+\frac{\lambda_2}{2}{\cal T}_{\rm E}[\bar G_{\rm E}]-\frac{\lambda^2}{2}\phi^2{\cal B}[\bar G_{\rm E}](Q_{\rm E})-\frac{\lambda^2}{6}{\cal S}_{\rm E}[\bar G_{\rm E}](Q_{\rm E})\,.
\label{Eq:ext_gap_mass1}
\eeq
We mention that, in the ${\cal O}(\lambda^2)$ truncation, the internal propagator has exactly the same expression as the approximate external propagator introduced above, since the gap mass $\bar M_{\rm E}^2(Q)$ is given by the right hand side of \eqref{Eq:ext_gap_mass1}. 

In fact, we can do a little bit better concerning the external propagator, as the zero-momentum value $\hat M_{\rm E}^2\equiv\hat M_{\rm E}^2(Q_{\rm E}=0)$ does not need to be approximated, but can be computed instead from curvature of the effective potential associated to the effective action. Proceeding this way, one completely takes into account the contribution of the Bethe-Salpeter equation to $\hat M_{\rm E}^2$ and away from $Q_{\rm E}=0$ we can approximate $\hat M_{\rm E}^2(Q_{\rm E})$ by
\beq
\hat{M}_{\rm E}^2(Q_{\rm E})=\delta Z_2 Q_{\rm E}^2+\hat M_{\rm E}^2-\frac{\lambda^2}{2}\phi^2\Big[{\cal B}[\bar G_{\rm E}](Q_{\rm E})-{\cal B}[\bar G_{\rm E}]\Big]-\frac{\lambda^2}{6}\Big[{\cal S}_{\rm E}[\bar G_{\rm E}](Q_{\rm E})-{\cal S}_{\rm E}[\bar G_{\rm E}]\Big]\,.
\label{Eq:ext_gap_mass_f}
\eeq
The counterterm $\delta Z_2$ is needed to absorb divergences in the difference of setting-sun integrals and is determined numerically as explained in Appendix A.2 of Ref.~\cite{Marko:2015gpa}, using instead of $T_\star$ a small value of the temperature, $T=0.05 T_\star.$ If we denote by $\hat M^2(Q)$ the analytic continuation of $\hat M^2_{\rm E}(Q_{\rm E})$, the pole mass of the external propagator corresponds to $\hat M^2_{\rm p}=\hat M^2(\hat M^2_{\rm p})$. 

The generalization of Eq.~\eqref{Eq:ext_gap_mass_f} to the $O(4)$ case is straightforward. To get the approximated $\hat M^2_{\rm L,E}(Q)$ and $\hat M^2_{\rm T,E}(Q)$ in the corresponding longitudinal and transverse external propagators, one can use the expressions given in Eq.~(15) of Ref.~\cite{Marko:2015gpa}, subtract from the right hand side the zero momentum expression and add the corresponding curvature mass determined numerically using  Eq.~(16) of that reference. We emphasize that the approximated expressions are the same in the two-loop and ${\cal O}(\lambda^2)$ truncations, but the internal propagators entering the integrals are different in the two cases.

\subsection{On the quality of the curvature mass based parametrization of the O(4)
model}

\begin{figure}
\begin{center}
\includegraphics[width=0.45\textwidth]{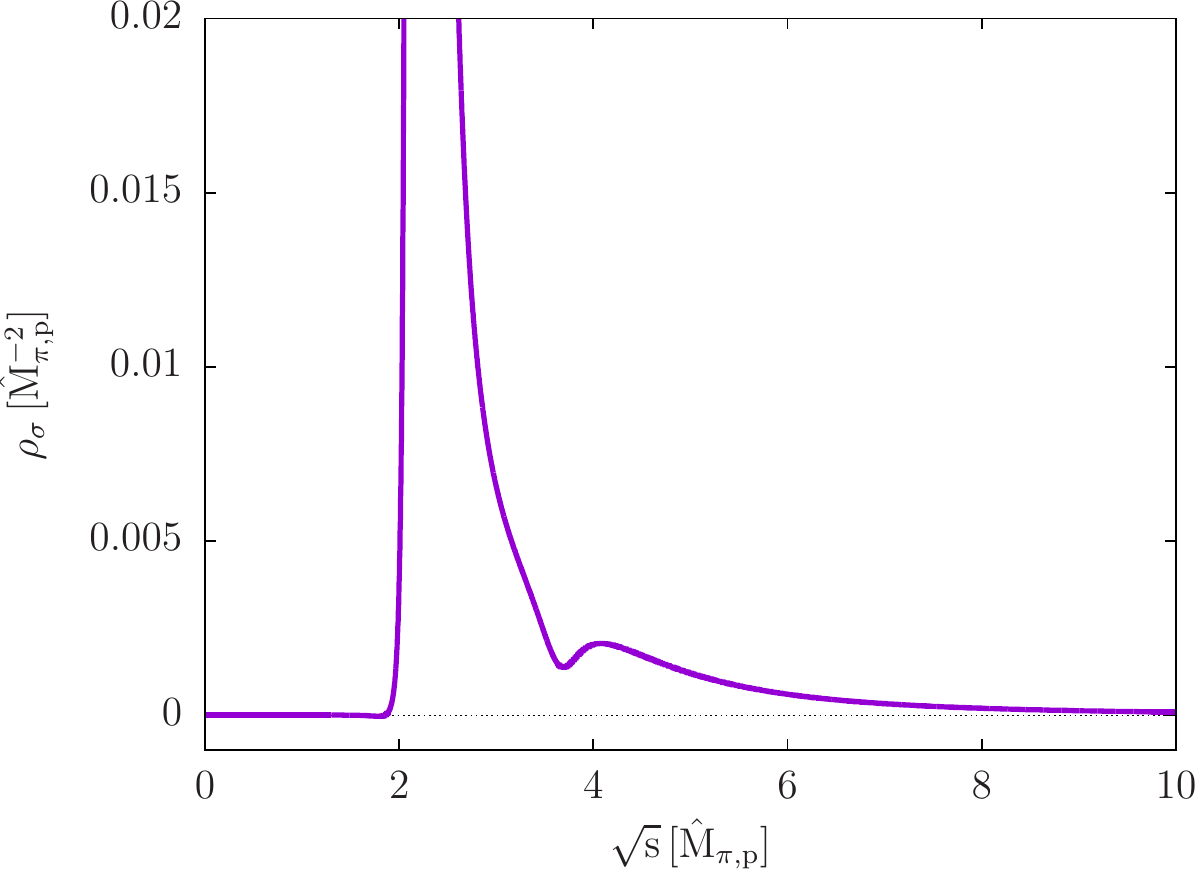}\hspace{0.05\textwidth}\includegraphics[width=0.45\textwidth]{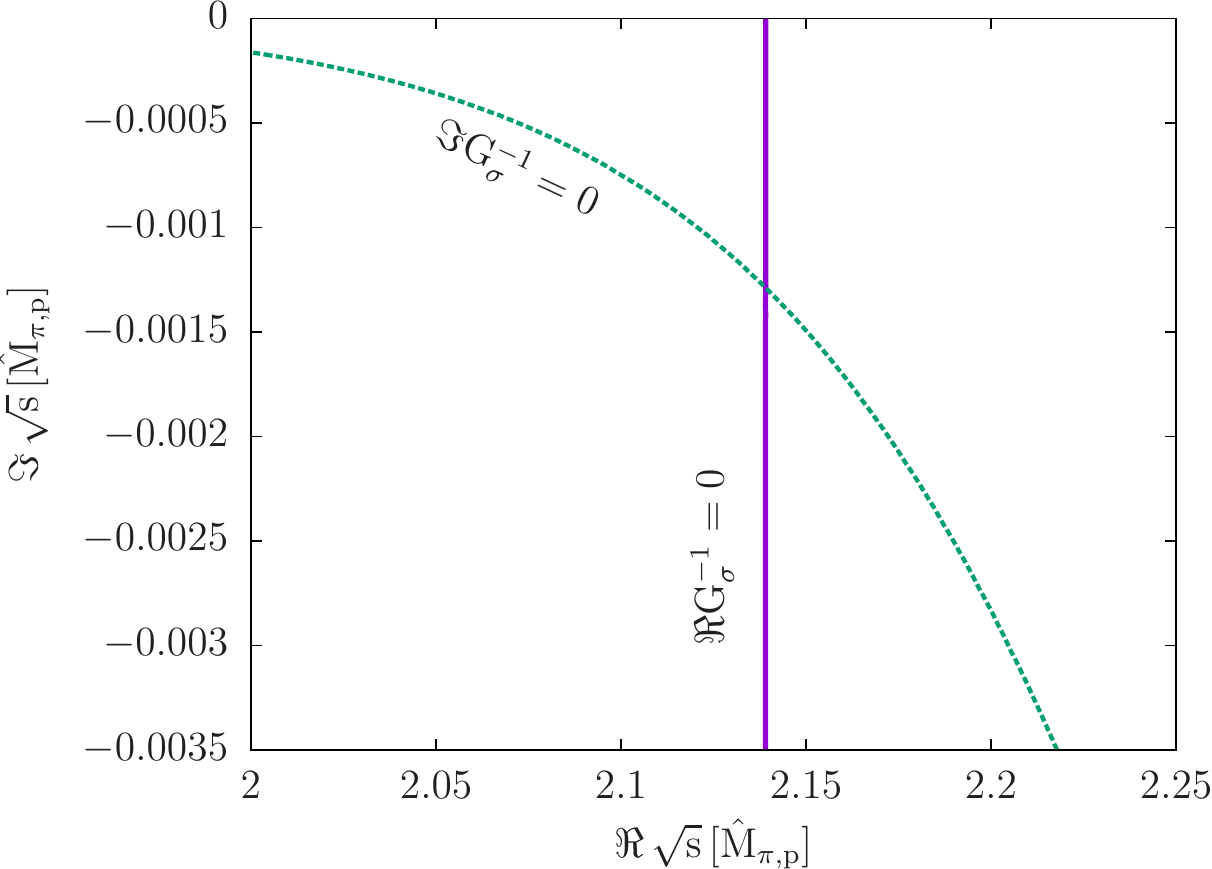}\caption{The result of the analytic continuation of Euclidean data for the external propagator obtained with a two-loop truncation for a typical parameter set ($m^2=0.23,\,\lambda=28.257,\,h=0.62$). Left panel: the spectral function in the sigma channel. Right panel: the zero-contour lines of the real and imaginary parts of the inverse $\sigma$ propagator showing the complex sigma pole with a very small imaginary part.\label{Fig:complexTS}}
\end{center}
\end{figure}

We now turn to the analytic continuation of the $N=4$ Euclidean results obtained in \cite{Marko:2013lxa,Marko:2015gpa} at a small value of the temperature, $T=0.05T_\star$. Applying the method based on Pad{\'e} approximants we find similar results in the two investigated truncations for any parameters where the curvature masses (zero-momentum masses of the external propagator) are physical. The main result is that although the difference between the pole masses of the longitudinal external and internal propagators can be of 40\%, the difference between the pole and zero momentum mass of a given propagator (internal or external) is only within 1\%. For the transverse mode the differences are typically smaller, which is in line with the results of \cite{Strodthoff:2016pxx, Helmboldt:2014iya} obtained using the FRG approach. We already knew from previous studies that the difference between $\hat M_{\rm L}$ and $\bar M_{\rm L}$ could be more than 30\% for some parameters (see Fig.~3 of \cite{Marko:2015gpa}) and now we see that both $\bar M_{\rm L/T}-\bar M_{\rm L/T,p}$ and $\hat M_{\rm L/T}-\hat M_{\rm L/T,p}$ are very small in both truncations, for parameters where the curvature masses are physical. For such parameters the pole of the analytically continued sigma propagator becomes complex. Based on the study presented in Sec.~III.B, we interpret this pole as being the physical pole on the 2nd Riemann sheet. Unfortunately, the imaginary part of the pole is almost zero: the ratio between real and imaginary part is $\sim 10^{-3}$. We illustrate this for a typical parameter set of the two-loop truncation in Fig.~\ref{Fig:complexTS}, where the plots are obtained by continuing the Euclidean external longitudinal propagator calculated with the approximation described above.

The smallness of the imaginary part of the complex $\sigma$ pole is in line with the findings of Ref.~\cite{Patkos:2002xb}, where it turned out that at leading order in a $1/N$ expansion its ratio to the real part of the pole only starts to grow for larger values of the coupling and becomes physically acceptable for $\lambda\in(300,400).$ We mention that taking into account the scaling by $N=4$ used in Refs.~\cite{Marko:2013lxa,Marko:2015gpa} the coupling of Ref.~\cite{Chiku:1998kd} (also used in Sec.~III.B) corresponds to $\lambda=292.$ Even for such a big value of the coupling constant $M_{\rm L}(Q=0)$ is only 1.7\% larger than $M_{\rm L,p}$ given in Sec.~III.B. Such a large value of the coupling, which is by a factor of 5 larger than the largest coupling used for parametrization in Refs.~\cite{Marko:2013lxa,Marko:2015gpa}, cannot be reached in our 2PI investigations because the closeness of the Landau pole to the physical scales makes the solution of the propagator equation, if accessible at all iteratively, highly cutoff sensitive.

%
\subsection{Infrared sensitivity of the pole mass in the two-loop SI2PI approximation\label{ss:SI2PI}}
\begin{figure}
\begin{center}
\includegraphics[width=0.45\textwidth]{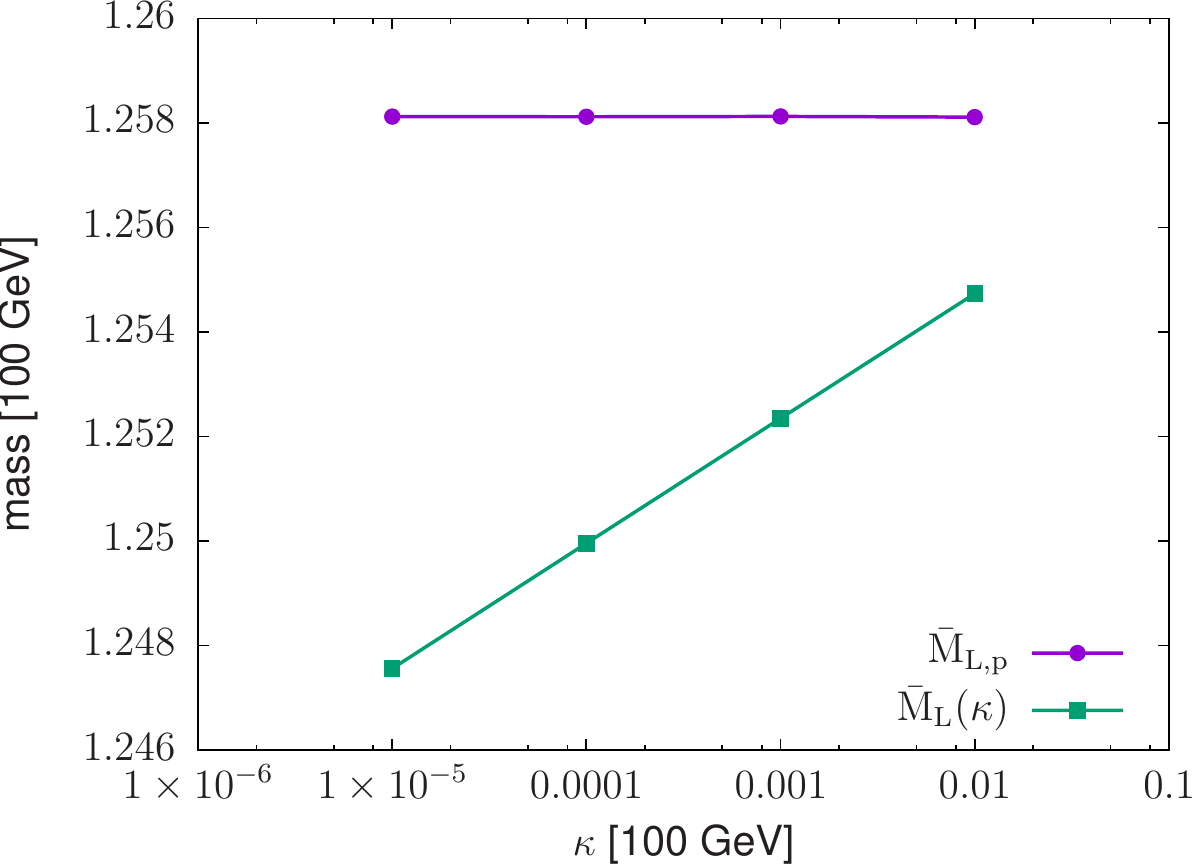}\hspace{0.05\textwidth}
\caption{Infrared sensitivity of the zero-momentum and pole ``Higgs'' masses in an $O(2)$-model using the symmetry improved 2PI framework. The solution is in principle lost if one insists in removing the infrared regulator $\kappa$ (which is inversely proportional to the typical linear size of the system). However before this happens, the pole mass presents a plateau behavior that allows to define the mass of the particle in a very accurate way. We used the parameters of Ref.~\cite{Pilaftsis:2013xna} and the non-equidistant grid of Ref.~\cite{Marko:2016wtw} with $N_s=151$ points. All these points in the $\sqrt{s}\in [\kappa,\Lambda]$ range, with $\Lambda=5$~TeV, were used to construct the Pad{\'e} approximant. \label{Fig:sI2PI} }
\end{center}
\end{figure}

In Ref.~\cite{Marko:2016wtw}, we investigated the symmetry improved 2PI (SI2PI) formalism of Ref.~\cite{Pilaftsis:2013xna} at two-loop order, using various types of UV regulators. In particular, we pointed out that, for generic smooth regulators, the solution for the internal propagator of the SI2PI framework possess an untamed infrared sensitivity in the broken phase that leads to a loss of solution for large enough volumes. This may not be a problem though in cases where the volume at which the solution disappears is many order of magnitudes higher than the physical volume of the system under study, and that the quantity under scrutiny presents a {\it plateau} behavior for a large range of volumes below the volume at which the solution is lost. In Ref.~\cite{Marko:2016wtw}, we tested this scenario on the zero-momentum mass of the Higgs propagator in the model of Ref.~\cite{Pilaftsis:2013xna} and we observed sensible changes with the volume as we also illustrate in Fig.~\ref{Fig:sI2PI}. However, as pointed out to us by D. Teresi, in the same model, the Higgs pole mass, should be (almost) insensitive to the volume. This insensitivity was reported in Ref.~\cite{Pilaftsis:2017enx} and it is expected based on the fact that the Higgs pole mass is not related to the zero-momentum behavior of the propagator, which is the only region where the propagator is infrared sensitive in our case. However, we note that it is not clear which type of UV regulator was used in Refs.~\cite{Pilaftsis:2013xna,Pilaftsis:2017enx}, so one could {\it a priori} think that the insensitivity could also originate from the use of a sharp regulator that tames artificially the IR sensitivity, as explained in Ref.~\cite{Marko:2016wtw}. We can now test these scenarios using the Pad{\'e} continuation technique with various types of UV regulators. The result for the Higgs pole mass is compared to the zero-momentum Higgs mass in Fig.~\ref{Fig:sI2PI}. Imposing the constraint of the SI2PI formalism as $\bar M_{\rm T}(|K_{\rm E}|=\kappa)=0$, we observe that, even with a smooth UV regulator, for which the internal Higgs propagator ceases to exist beyond some volume, the corresponding pole shows a plateau behavior over a large range of volumes.

\section{Conclusions}

We searched for broken phase Minkowski space solutions in the $(m^2,\lambda)$ parameter space of the one-component $\varphi^4$ model at zero temperature using the 2PI effective action truncated at two-loop level. Since the Euclidean solution can be obtained from the spectral function, we could use the Minkowskian solution as a benchmark for the analytic continuation of the Euclidean propagator with the method based one multipoint Pad{\'e} approximants. The various tests we performed show that at zero temperature the Pad{\'e} analytic continuation is an efficient tool to obtain not only the spectral function, but also the complex pole of the propagator on the second Riemann sheet. However, it is expected that as the temperature grows the applicability of the method is limited by the decreasing number of available Matsubara modes in a given frequency interval, as we have argued in Sec.~\ref{ss:test1}.

We applied the Pad{\'e} analytic continuation on previously determined practically zero temperature Euclidean propagators of the $O(4)$ model and studied the relation between the pole and the gap mass at vanishing external momentum of both the internal and external 2PI propagators. We have searched for real poles of the transverse propagator and for complex poles of the longitudinal propagator on the 2nd Riemann sheet. For parameters where the longitudinal curvature mass (gap mass at vanishing external momentum of the external propagator) is physical, the real part of the corresponding pole is within 1\% of the curvature mass, both at two-loop and at ${\cal O}(\lambda^2)$ level truncations of the 2PI effective action. This shows that the error of the parametrization of \cite{Marko:2013lxa} and \cite{Marko:2015gpa}, where the curvature masses where used instead of the pole masses, is smaller than 1\%. However, with the parameters determined in these references the imaginary part of the pole of the longitudinal external propagator turns out to be very small, although it increases with the value of the coupling constant. Unfortunately, the presence of the Landau pole prevents us from accessing regions of the parameter space with larger values of the coupling, as the solution of the model, which is increasingly difficult to find with an iterative method, becomes highly cutoff sensitive.

\acknowledgments{We would like to thank D. Teresi for clarifying discussions about his work that motivated the analysis presented in Sec.~\ref{ss:SI2PI}. In case of G.~M.~ this work is part of Project no.~121064 for which support was provided from the National Research, Development and Innovation Fund of Hungary, financed under the PD\_16 funding scheme. Zs. Sz. would like to thank {\'E}cole Polytechnique for hospitality and support during the last stages of this work.}

\appendix

\section{Renormalization \label{sec:renorm}}

As mentioned in the Introduction above Eq.~\eqref{Eq:gap-field_F}, we now discuss how to obtain the counterterms with a renormalization at $T=0,$ in a way that resembles the renormalization prescriptions at $T=T_\star$ which were used in some of our previous works. Since there is no physical renormalization condition giving exactly the counterterms determined in \cite{Patkos:2008ik}, we define $m_0^2$ and $m_2^2$ by absorbing those quadratic and logarithmic divergences which arise when expanding the propagator $\bar G$ around $G_0(Q)=i/(Q^2-M_0^2+i\varepsilon)$ and do not depend on the environment ($\phi$ or ${\cal T}_{\rm F}[\bar G]$). One has
\beq
m_0^2&=&m^2-\frac{\lambda_0}{2}{\cal T}_{\rm div}[\bar G],\\
m_2^2&=&m^2-\frac{\lambda_0}{2}{\cal T}_{\rm div}[\bar G] - \frac{\lambda^2}{6}{\cal S}_{\rm div}[\bar G],
\eeq
where the environmental free divergences of the tadpole and setting-sun integrals are given by
\bea
{\cal T}_{\rm div}[\bar G] &=& {\cal T}[G_0] + (m^2-M_0^2) {\cal I}[G_0],\\
{\cal S}_{\rm div}[\bar G] &=& {\cal S}[G_0] + 3(m^2-M_0^2) \big[T_{\rm d}^{(I)}+{\cal I}^2[G_0]\big],
\eea
with $T_{\rm d}^{(I)}=-i\int_Q G_0^2(Q) {\cal I}_{\rm F}[G_0](Q).$

In order to renormalize the couplings we proceed as usual at $\phi=0$ (see Ref.~\cite{Berges:2005hc} for details) and introduce the 2PI kernels and the related four-point functions, but formally replace in them and in the equations they satisfy $\bar G_{\phi=0}$ by $G_0.$ One has for instance
\bea
\bar V_0 &=& \bar \Lambda_0+\frac{1}{2} \bar V_0 \bar\Lambda_0 {\cal I}[G_0],\\
V_0(0,K) &=&\Lambda_0(0,K)-\frac{i}{2}\int_Q\Lambda_0(0,Q) G_0^2(Q)\bar V_0,
\eea
where at the present order of truncation
\bea
\bar \Lambda_0&\equiv& 4 \left.\frac{\delta^2 \gamma_{\rm int}}{\delta G \delta G}\right|_{\phi=0,G_{\phi=0}\to G_0}=\lambda_0, \\
\Lambda_0(0,K)&\equiv& 2 \left.\frac{\delta^3 \gamma_{\rm int}}{\delta\phi\delta\phi\delta G(K)}\right|_{\phi=0,G_{\phi=0}\to G_0}=\lambda_2+\lambda^2{\cal I}[G_0](K).
\eea
$\lambda_0$ is obtained from the condition $\bar V_0=\lambda$ in the form $\lambda_0^{-1}=\lambda^{-1}+{\cal I}[G_0]/2$, while $\lambda_2$ from the condition $V_0(0,K)=\lambda,$ which gives after a few lines of algebra $\lambda_2=\lambda_0-\lambda^2 {\cal I}[G_0]-\lambda^2\lambda_0 T_{\rm d}^{(I)}/2.$ The counterterm $\delta \lambda_4$ is obtained from the condition $\hat V_0=\lambda,$ using the equation for
\beq
\hat V_0\equiv\left.\frac{\delta^4 \gamma}{\delta\phi^4}\right|_{\phi=0,G_{\phi=0}\to G_0}\,.
\eeq

\section{Dispersion relations and other useful relations}\label{app:disp}

Let us start recalling here Eqs.~(\ref{eq:spectral-rep}) and (\ref{Eq:sumrule}) can be understood from the sole assumption of the existence of an analytic propagator ${\cal G}(Q^2)$ for complex values of $Q^2$ away from the positive real axis (in the first Riemann sheet), that decreases as $1/Q^2$ at large $Q^2$ and which gives back the Minkowski propagator when approaching the positive real axis: $G(Q)={\cal G}(Q^2+i\epsilon)$ with $Q^2>0$. Indeed, the propagator ${\cal G}(Q^2)$ being analytic away from the positive real axis, one can apply Cauchy formula to a small circle not intersecting this axis and centered around a certain $Q^2$. Deforming this contour into a large circle $C_R$ whose radius $R$ tends to infinity and a contour $B_R$ with two parallel branches on each side and close to the positive real axis, one obtains
\be
{\cal G}(Q^2)=\int_{C_R}\frac{d s}{2\pi i}\,\frac{{\cal G}(s)}{s-Q^2}+\int_0^R \frac{d s}{2\pi i}\frac{{\cal G}(s+i\epsilon)-{\cal G}(s-i\epsilon)}{s-Q^2}\,.
\ee
Owing to the assumed behavior of ${\cal G}(s)$ at large $s$, the first term goes to $0$ as $R\to\infty$. The second term can be expressed in terms of the spectral function and we finally obtain
\be\label{eq:analytic}
{\cal G}(Q)\equiv\int_0^\infty \frac{d s}{2\pi}\frac{i\rho(s)}{Q^2-s}\,,
\ee
which provides a spectral representation for the analytic propagator that leads to Eq.~(\ref{eq:spectral-rep}) when approaching the positive real axis. The sum rule is obtained in a similar fashion by applying the argument to the function $(s-Q^2){\cal G}(s)$ instead of ${\cal G}(s)$. In this case the integral over the contour $C_R$ does not vanish and gives the $1$ in the right-hand-side of Eq.~(\ref{Eq:sumrule}). 

The previous argument can be repeated for any function ${\cal F}(Q^2)$ analytic in $Q^2$ away from the positive real axis. In the case where ${\cal F}(s)$ does not go to zero fast enough, one should apply the argument to ${\cal F}(s)-\sum_{p=0}^n {\cal F}^{(p)}(s_0) (s-s_0)^p/p!$ with $n$ high enough, such that the integral over $C_R$ vanishes as $R\to\infty$. For instance, if ${\cal F}(s)$ grows logarithmically, we apply the argument to ${\cal F}(s)-{\cal F}(s_0)$ with $s_0=Q_0^2$ away from the positive real axis. We obtain, for any $Q^2$ and $Q_0^2$ away from the positive real axis,
\be
{\cal F}(Q^2)-{\cal F}(Q_0^2)=\int_0^\infty\frac{d s}{\pi}\frac{(Q^2-Q_0^2)\Im F(s)}{(s-Q^2)(s-Q_0^2)}\,.
\ee
In particular, if one is interested in $F(Q)\equiv {\cal F}(Q^2+i\epsilon)$, one has for $Q^2$ and $Q_0^2$ real,
\be
F(Q)-F(Q_0)=\int_0^\infty\frac{d s}{\pi}\frac{(Q^2-Q_0^2)\Im F(s)}{(s-Q^2-i\epsilon)(s-Q_0^2-i\epsilon)}\,.
\ee
The previous formula applies in particular to the finite bubble integral considered in the present work. Indeed, using the spectral representation and its sum rule,
\beq
{\cal I}_F[G](Q)=\int_0^\infty\frac{d s_1}{2\pi}\int_0^\infty\frac{d s_2}{2\pi}\rho(s_1)\rho(s_2){\cal I}_{0,\rm F}[G_1,G_2](Q)\,,
\eeq
from which it follows that this is analytic in the variable $Q^2$ away from the positive real axis. Moreover since the finite bubble grows like $\ln Q^2$, we need to apply the once-subtracted formula.\\

Let us now derive some other useful results \cite{Sauli:2004bx}. We start from the identity $\bar G(Q) \bar G^{-1}(Q)=1$ and derive two expressions for the finite wave-function renormalization constant $Z,$ the expression for the continuum part of the spectral function and the sum rule it satisfies. For $\bar G(Q)$ we use its spectral representation \eqref{eq:spectral-rep}, while for the inverse propagator we use the expression $\bar G^{-1}(Q)=-i(Q^2-\bar M^2(Q)+i\varepsilon).$ The gap mass is a complex valued function and $\Re\bar M^2(Q)$ and $\Im\bar M^2(Q)$ can be read off from \eqref{Eq:gap_F}, remembering that only the bubble integral has imaginary part. Using the form of the spectral function given in \eqref{eq:rhoPolesep} and the relation
\be
\frac{1}{Q^2 - s + i\varepsilon}={\cal P}\frac{1}{Q^2-s}-i\pi\delta(Q^2 - s)\,,
\ee
one obtains from $G G^{-1}=1$ two equations, one for the real part and one for the imaginary part. With some algebraic manipulations one derives from them the usual expression for the continuum part of the spectral function
\beq
\label{eq:sigma}
\sigma(Q^2) = -2\frac{\Im\bar M^2(Q)}{(Q^2-\Re\bar M^2(Q))^2+(\Im\bar M^2(Q))^2}\,,
\eeq
and the equation ($s_{\rm th}=4 \bar M^2_{\rm p}$)
\beq
Z+(Q^2-\bar M^2_{\rm p})\,{\cal P}\int_{s_{\rm th}}^\infty\frac{d s}{2\pi}\frac{\sigma(s)}{Q^2-s}=\frac{(Q^2-\bar M^2_{\rm p})(Q^2-\Re\bar M^2(Q))^2}{(Q^2-\Re\bar M^2(Q))^2+(\Im\bar M^2(Q))^2}.
\label{Eq:rel_from_GGinv}
\eeq
Taking the limit $Q\to \infty$ in \eqref{Eq:rel_from_GGinv} one obtains
\beq
Z+\int_{s_{\rm th}}^\infty\frac{d s}{2\pi}\sigma(s) = 1,
\label{eq:sumrule_split}
\eeq
which is the sum rule for a spectral function of the form \eqref{eq:rhoPolesep}, while taking the limit $Q\to \bar M_{\rm p},$ one obtains
\beq
Z=\left(1-\frac{d\,\Re\bar M^2(Q)}{d Q}\bigg|_{Q=\bar M_{\rm p}}\right)^{-1}\,.
\label{eq:residue}
\eeq
which is the residue of the pole $Q=\bar M_{\rm p}$ of the propagator.

\section{Numerical algorithm and its implementation}\label{app:algo}
In this Appendix we describe the steps of the iterative process applied to solve sequentially the explicitly finite coupled field and gap equations \eqref{Eq:gap-field_F} and present the numerical implementation of these steps. 

Being interested in broken symmetry phase solutions, we start from an initial propagator $\bar G^{(0)}$ and express the nontrivial $(\bar\phi^2)^{(1)}$ from the field equation. Then, using both $\bar\phi^{(1)}$ and $\bar G^{(0)},$ we evaluate $(\bar M^2)^{(1)}$ from \eqref{Eq:gap_F}, which in turn gives us $\bar G^{(1)}(Q)$. These steps are repeated until the relative change from one iteration to the next of both $\bar\phi$ and the pole mass $\bar M^2_{\rm p}$ obtained from \eqref{eq:poleEq} is less than our stopping parameter, chosen to be $10^{-7}.$ If $(\bar\phi^2)^{(i+1)}<0$ we leave $\bar\phi^{(i+1)}=\bar\phi^{(i)}$. When for 5 consecutive iterative steps $\bar\phi^2$ is negative, we stop the iteration and say that no broken phase solution was found for the given parameters and/or initial values. 

The explicit steps for one full iteration (from the $i$-th to $i+1$-th) are the following. The propagator $G^{(i)}$ is used to denote a given quantity in the $i$th iteration, even though it is clear from Sec.~\ref{ss:disp_rel} that the spectral function of the form \eqref{eq:rhoPolesep} is the central object of the entire iterative procedure.
\begin{enumerate}
\item Using \eqref{Eq:TIS_F} and \eqref{Eq:dG_Gr_Ml}, evaluate ${\cal S}_{\rm F}[\bar G_{\rm E}^{(i)}]$ and ${\cal T}_{\rm F}[\bar G_{\rm E}^{(i)}]$ and update $\bar\phi^{(i+1)}$ from \eqref{Eq:field_F}. 
\item Evaluate $\Im {\cal I}[\bar G^{(i)}](Q)$ using \eqref{eq:ImG} and then $\Re {\cal I}_{\rm F}[\bar G^{(i)}](Q)$ using the dispersion relation \eqref{Eq:dispRel}.
\item Determine $(\bar M^2_{\rm p})^{(i+1)}$ solving the pole equation \eqref{eq:poleEq} and update $\sigma^{(i+1)}(s)$ using its expression given in \eqref{eq:sigma}, then update $Z$ using the sum rule \eqref{eq:sumrule_split}.\footnote{Note that by changing $\bar M^2_{\rm p}$ in each iteration $s_{\rm th}$ also changes. Due to the structure of the bubble integral $(s_{\rm th})^{(i+1)}=4(\bar M^2_{\rm p})^{(i)}$.} 
\end{enumerate}

We initialize $\bar\phi$ with some crude non-zero estimate of the solution ({\it e.g.} the classical value $\sqrt{-m^2/\lambda}$) and choose the initial spectral function to contain only a pole part with unit residue, with the pole at $\bar M_{\rm H}$, defined as the solution of the gap equation in the Hartree approximation
\beq
\bar M_{\rm H}^2 = m^2+\frac{\lambda}{2}\big(\bar\phi^2+{\cal T}_{\rm F}[\bar G_{\rm H}]\big)\,,
\label{Eq:M2_H}
\eeq
with $\bar G_{\rm H}(Q) = i/(Q^2-\bar M_{\rm H}^2+i\varepsilon)$.\\

We note that one can introduce a reparametrization of the equations \eqref{Eq:gap-field_F} in terms of the mass parameter
\beq
\label{Eq:M2}
M^2=\bar M^2_{\phi=\bar\phi}(Q=0)\,.
\eeq
that is the zero momentum value of the self-energy. While this introduces some intricacy connected to the fact that $M^2$ depends on the {\it a priori} unknown solution $\bar\phi,$ it has several advantages. First, by fixing the local part of the self-energy, the number of iterations required for convergence is greatly reduced. Also, the tadpole diagram \eqref{Eq:tad}, one of the main sources of numerical error, only has to be evaluated to obtain $m^2$ corresponding to the fixed value of $M^2$. Second, this allows the introduction of an underrelaxation parameter $\alpha$ as
\beq
\label{eq:alpha}
(\bar\phi_{\alpha}^{(i+1)})^{2}=\alpha(\bar\phi^{(i+1)})^{2}+(1-\alpha)(\bar\phi^{(i)})^{2}\,.
\eeq
This is needed to achieve convergence in parameter regions where otherwise our iterative method fails. This combined with the good initial guess provided by $M^{2}$ allowed us to find the unphysical solution shown in Fig.~\ref{Fig:nonph}.

\subsection{Numerical implementation}\label{sec:numImp}

In the iterative method described above we have to store $\bar\phi,\,\bar M^2_{\rm p},$ and discretized version of $\sigma(s)$. For the latter we use a grid of $N_s$ points (typically we use $N_s=500$) defined actually in the transformed variable $t=\frac{s}{1+s}$. Starting from $t(s=s_{\rm th})$ and ending in $1$ one has 
\beq
\label{eq:tGrid}
t_i = \kappa+i^3\times\Delta\,,\quad i=0\dots N_s-1\,,
\eeq
with $\kappa=\frac{s_{\rm th}}{s_{\rm th}+1}$ and $\Delta=\frac{1-\kappa}{(N_s-1)^3}$. The property $t(s=\infty)=1$ allows us to carry out the $s$-integrals of Sec.~\ref{ss:disp_rel} without introducing an explicit numerical cutoff.\\

We follow below the order of the iterative steps described above and give some technical details concerning the evaluation of the integrals.\\

The explicitly finite expression of the two local Euclidean integrals, the tadpole ${\cal T}_{\rm F}[\bar G_{\rm E}]$ and ${\cal S}_{\rm F}[\bar G_{\rm E}]$ is given in Appendix~\ref{sec:integrals}, in \eqref{eq:tadFexpl} and (\ref{eq:SSGrexpl})--(\ref{eq:SSRestexpl}) respectively, in terms of the Euclidean propagator \eqref{Eq:spectral-rep_E}. The integral of \eqref{Eq:spectral-rep_E} is evaluated on a rather complicated looking momentum grid defined as
\beq
q_i=i^{\alpha(i)}\times \Delta(i)\,,\quad \alpha(i)=2+\frac{2}{\pi}\,{\rm atan}\,\left(\,{\rm sinh}\,\left(5-\frac{10i}{N_s-1}\right)\right)\,,\quad i=0\dots N_s-1\,,
\eeq
with $\Delta(i) = \Lambda/(N_s-1)^{\alpha(i)},$ where $\Lambda$ is an appropriate numerical cutoff for which we typically use values in the range $(50-200).$\footnote{In principle we could also use a transformed variable similar to \eqref{eq:tGrid} and integrate up to infinity since the integrals are finite. However, the integrands of \eqref{eq:tadFexpl} and (\ref{eq:SSGrexpl})--(\ref{eq:SSRestexpl}) only decrease fast enough due to cancellations, which break down when numerical number representation errors are comparable to the results.} The reason for using such a grid is to sample with sufficient accuracy the Euclidean propagator both in the IR and UV regions. For integration we need to know $G_{\rm E}$ in between the grid points, but in order to improve on the UV behavior, we subtract $1/(q_i^2+M_0^2)$ from $G_{\rm E}(q_i)$ and fit the difference with a Steffen-spline.\footnote{This is used because Steffen's method leads to an interpolation function which is monotonic between the given data points.} This proves to be accurate enough in the UV to see the apparent convergence of the integrals with increasing $\Lambda$. The spline is evaluated in the integrand called by the {\it cquad} adaptive integration routine of GSL library \cite{GSL}. \\

To compute $\Im {\cal I}[\bar G](Q),$ we use \eqref{eq:rhoPolesep} in \eqref{eq:ImG}. This leads to $\delta\times\delta$, $\delta\times\sigma$, and $\sigma\times\sigma$ type terms, which start giving non-vanishing contribution for $Q>2\bar M_{\rm p},$ $Q>3\bar M_{\rm p},$ and $Q>4\bar M_{\rm p},$ respectively. Since $\Im {\cal I}[\bar G](Q)$ will be used in the dispersion relation \eqref{Eq:dispRel} giving $\Re {\cal I}_{\rm F}[\bar G](Q)$, we spline it in order to be able to evaluate it anywhere. We use the grid introduced for $\sigma$ in \eqref{eq:tGrid} in the same transformed coordinates. However, a spline interpolation cannot reproduce a functional form with infinite derivatives, such as $\Im {\cal I}[\bar G](Q),$ which behaves as a square root function around $Q=2\bar M_{\rm p}.$ Therefore, in order to avoid this problem, we chose to fit with a cubic spline the function $\left(\Im {\cal I}[\bar G](Q)\right)^2.$\\

$\Re {\cal I}_{\rm F}[\bar G](Q)$ is evaluated from the dispersion relation \eqref{Eq:dispRel}. In order to circumvent the problem of numerically computing integrals with a principal value prescription when $s=Q^2>4\bar M^2_{\rm p},$ we rewrite them with the usual method (see {\it e.g.} \cite{Shima}). First we split the interval of integration by introducing the point $2s-4\bar M^2_{\rm p},$ then we add and subtract an appropriate term in the integrand of the resulting integral over the interval $[4\bar M^2_{\rm p},2s-4\bar M^2_{\rm p}],$ exploiting the fact that ${\cal P} \int_{a}^{2b-a} d x/(x-b)=0$. In this way we obtain
\beq
{\cal P}\int_{4\bar M^2_{\rm p}}^\infty d s'\frac{\Im {\cal I}[\bar G]\big(\sqrt{s'}\big)}{s'(s'-s)} = {\cal P}\int_{4\bar M^2_{\rm p}}^{2s-4\bar M^2_{\rm p}}\frac{d s'}{s'-s}\left(\frac{\Im {\cal I}[\bar G]\big(\sqrt{s'}\big)}{s'}-\frac{\Im {\cal I}[\bar G]\big(\sqrt{s}\big)}{s}\right)+\int_{2s-4\bar M^2_{\rm p}}^\infty d s'\frac{\Im {\cal I}[\bar G]\big(\sqrt{s'}\big)}{s'(s'-s)}\,.
\label{Eq:dispRel2}
\eeq
Actually, the integrand of the first integral on the r.h.s. is continuous at $s'=s,$ so that the principal value prescription can be omitted when integrating it numerically with the {\it cquad} routine. \\

We end this part by mentioning two numerical intricacies. The first is encountered when fitting $\sigma$ with a spline on the grid \eqref{eq:tGrid}. $\sigma$ grows out at the threshold as a square root function, similarly to $\Im{\cal I}[\bar G]$. However, contrary to $\Im{\cal I}[\bar G]$, which enters integrals which decrease fast enough in the UV (see the suppression by powers of $s$ in \eqref{Eq:dispRel2}), the convergence of the integrals with $\sigma$ (see {\it e.g.} \eqref{eq:ImG} and \eqref{eq:sumrule_split}) is assured by the UV behavior of $\sigma$, which has to be preserved by the fit. This time, fitting $\sigma^2$ would not be enough because when computing integrals using the transformed variable $t$ (see the definition in the paragraph before \eqref{eq:tGrid}) the Jacobian $J(t)=(1-t)^{-2}$ of the transformation is divergent as $t\to1.$ This would lead in the integrals involving $\sigma$ to a blowing-up of the oscillations of the spline when $t\to1.$ Since $\underset{t\to1}{\lim}\,J(t)\sigma(t)$ is finite, the way out is to fit the function $(J(t)\sigma(t))^2$ with a spline on the grid introduced in \eqref{eq:tGrid}. In this way the spline respects the requirements of having appropriate behaviors around the threshold and in the UV.

\begin{figure}
\begin{center}
\includegraphics[width=0.485\textwidth]{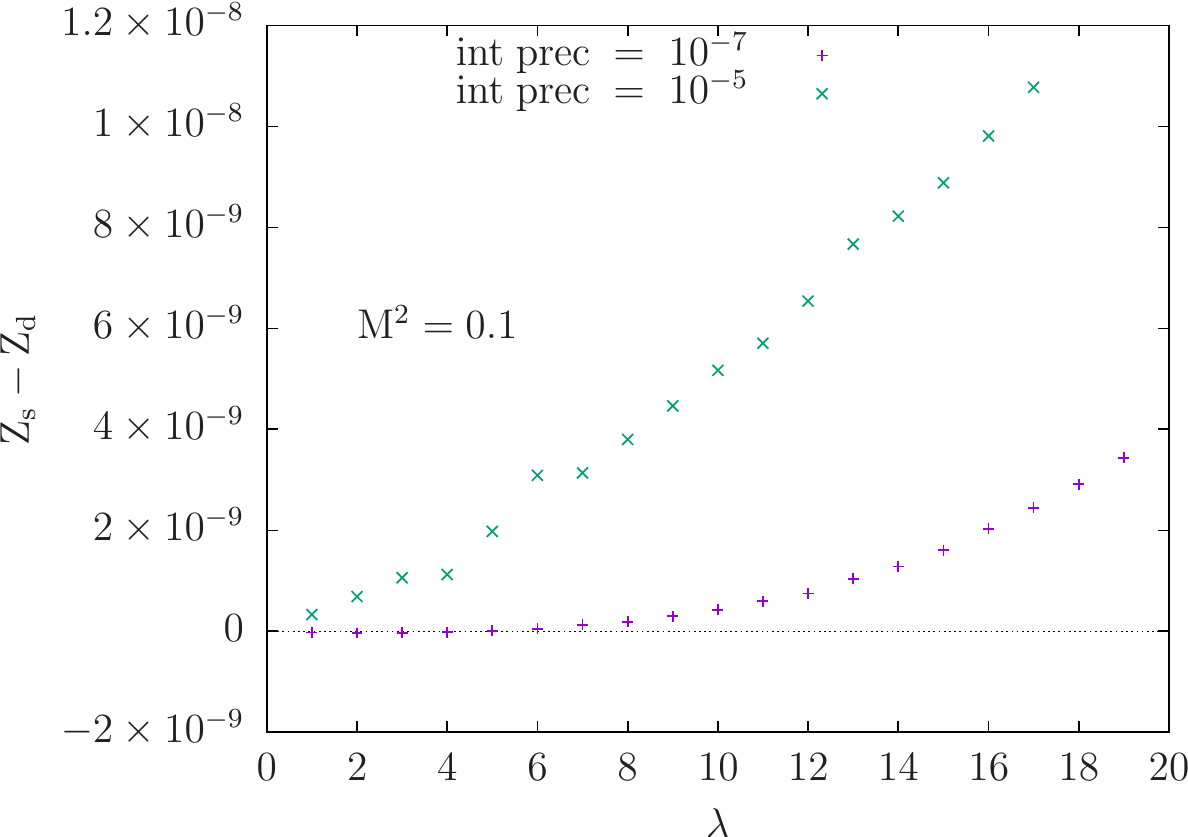}
\caption{We show, as a function of the coupling $\lambda$, the difference between the two ways of determining the coefficient $Z$ of the singular, pole part of the spectral function: when calculated from \eqref{eq:residue} as the residue of the pole it is denoted by $Z_{\rm d}$ and when obtained from the sum rule \eqref{eq:sumrule_split}, as done during the iterations, it is denoted by $Z_{\rm s}.$ See the text for more details.\label{fig:Zdiff}}
\end{center}
\end{figure}

The second intricacy is related to the coefficient $Z$ of the pole part of the spectral function. This has to be evaluated from iteration to iteration either from the sum rule \eqref{eq:sumrule_split} or as the residue of the pole \eqref{eq:residue}. While in principle the two ways of computing $Z$ are equivalent, numerical differences could occur, as we show in Fig.~\ref{fig:Zdiff}. There we present the value of $Z$ computed in the above mentioned two ways at the end of the iterative method. It turned out that for the iterations to converge, it is actually very important to ensure that the sum rule is satisfied iteration by iteration, since this guarantees the proper asymptotic behavior of the propagator, which was also used to renormalize the real part of the bubble (see Sec.~\ref{ss:disp_rel}). Hence, when solving the equations, $Z$ is always evaluated from the sum rule \eqref{eq:sumrule_split}. The difference seen in Fig.~\ref{fig:Zdiff} has purely a numerical reason. Increasing the precision of the numerical integration substantially reduces the difference which grows with increasing $\lambda.$

\section{Explicit form of some integrals \label{sec:integrals}}

In this Appendix we give the expression of the finite Euclidean tadpole \eqref{Eq:tad_F} and setting-sun \eqref{Eq:SS_F} integrals after analytically performing as many of their angular integrals as possible. The cutoff regularization used is such that an integral involving at least two propagators is invariant with respect to shifts of the loop momenta (see {\it e.g.} Appendix~D of \cite{Reinosa:2011cs}). The parts containing $G_{\rm r}$, that is \eqref{Eq:tad_F} and the third term of \eqref{Eq:SS_F}, reads ($q=|Q_{\rm E}|$)
\beq
{\cal T}_{\rm F}[G_{\rm E}]=\frac{1}{8\pi^{2}}\int_0^{\Lambda}d q\,q^3 G_{\rm r,E}(q)\,,\label{eq:tadFexpl}
\eeq
\beq
\int_{Q_{\rm E}}G_{\rm r,E}(Q_{\rm E}){\cal B}_{\rm F}[G_{\rm 0,E}](Q_{\rm E})=\frac{1}{8\pi^{2}}\int_0^{\Lambda}d q\,q^3 {\cal B}_{\rm F}[G_{\rm 0,E}](q)G_{\rm r,E}(q)\,,\label{eq:SSGrexpl}
\eeq
where
\[
G_{\rm r,E}(q)=\delta G_{\rm E}(q)+\left(M_{\rm l}^2-M_0^{2}-\frac{\lambda^2}{2}\phi^2{\cal B}_{\rm F}[G_{\rm 0,E}](q)\right)\,G^2_{\rm 0,E}(q)\,,
\]
with $\delta G_{\rm E}(q)=G_{\rm E}(q)-G_{\rm 0,E}(q)$ and $G_{\rm 0,E}(q)=(q^2+M_0^2)^{-1}.$ Observing that the other two terms of \eqref{Eq:SS_F} can be combined, one obtains 
\beq
\nonumber {\cal S}_{\rm E}[\delta G_{\rm E}]+3{\cal S}_{\rm E}[\delta G_{\rm E};\delta G_{\rm E};G_{\rm 0,E}]&=&\frac{1}{64\pi^5}\int_0^\Lambda d k\,k\,\delta G_{\rm E}(k)\int_0^\Lambda d p\,p\, \delta G_{\rm E}(p)\\
&&\times\int_{|k-p|}^{{\rm min}(\Lambda,k+p)}d q\,q\sqrt{-\lambda(q^2,k^2,p^2)}\big(G_{\rm E}(q)+2 G_{\rm 0,E}(q)\big)\,,\label{eq:SSRestexpl}
\eeq
with $\lambda(q^2,k^2,p^2)$ defined after \eqref{eq:ImG}.


\begin{thebibliography}{99}

\bibitem{Braun:2007bx} 
  J.~Braun, H.~Gies and J.~M.~Pawlowski,
  Phys.\ Lett.\ B {\bf 684}, 262 (2010).
  
\bibitem{Braun:2014ata} 
  J.~Braun, L.~Fister, J.~M.~Pawlowski and F.~Rennecke,
  Phys.\ Rev.\ D {\bf 94}, 034016 (2016).
  
\bibitem{Contant:2017gtz} 
  R.~Contant and M.~Q.~Huber,
  arXiv:1706.00943 [hep-ph].
  
\bibitem{Reinhardt:2017pyr} 
  H.~Reinhardt, G.~Burgio, D.~Campagnari, E.~Ebadati, J.~Heffner, M.~Quandt, P.~Vastag and H.~Vogt,
  arXiv:1706.02702 [hep-th].
  
\bibitem{Sauli:2001mb} 
  V.~\v Sauli and J.~Adam,
  Nucl.\ Phys.\ A {\bf 689}, 467 (2001).

\bibitem{VanHees:2001pf} 
  H.~Van Hees and J.~Knoll,
  Phys.\ Rev.\ D {\bf 65}, 105005 (2002).

\bibitem{Cooper:2004rs} 
  F.~Cooper, B.~Mihaila and J.~F.~Dawson,
  Phys.\ Rev.\ D {\bf 70}, 105008 (2004).

\bibitem{Sauli:2004bx} 
  V.~\v Sauli,
  Few Body Syst.\ {\bf 39}, 45 (2006).

\bibitem{Arrizabalaga:2007zz} 
  A.~Arrizabalaga and U.~Reinosa,
  Eur.\ Phys.\ J.\ A {\bf 31}, 754 (2007).

\bibitem{Jakovac:2006dj} 
  A.~Jakov\'ac,
  Phys.\ Rev.\ D {\bf 74}, 085026 (2006).

\bibitem{Jakovac:2006gi} 
  A.~Jakov\'ac,
  Phys.\ Rev.\ D {\bf 76}, 125004 (2007).

\bibitem{Kamikado:2013sia} 
  K.~Kamikado, N.~Strodthoff, L.~von Smekal and J.~Wambach,
  Eur.\ Phys.\ J.\ C {\bf 74}, 2806 (2014).

\bibitem{Strodthoff:2016pxx} 
  N.~Strodthoff,
  Phys.\ Rev.\ D {\bf 95}, 076002 (2017).

\bibitem{Jarrell:1996rrw} 
  M.~Jarrell and J.~E.~Gubernatis,
  Phys.\ Rept.\  {\bf 269}, 133 (1996).

\bibitem{Burnier:2013nla} 
  Y.~Burnier and A.~Rothkopf,
  Phys.\ Rev.\ Lett.\  {\bf 111}, 182003 (2013).

\bibitem{Kim:2014iga} 
  S.~Kim, P.~Petreczky and A.~Rothkopf,
  Phys.\ Rev.\ D {\bf 91}, 054511 (2015).

\bibitem{Fischer:2017kbq} 
  C.~S.~Fischer, J.~M.~Pawlowski, A.~Rothkopf and C.~A.~Welzbacher,
  arXiv:1705.03207 [hep-ph].

\bibitem{Carrington:2013jta} 
  M.~E.~Carrington, W.~J.~Fu, P.~Mikula and D.~Pickering,
  Phys.\ Rev.\ D {\bf 89}, 025013 (2014).
  
\bibitem{VS_Pade}
  H.~J.~Vidberg and J.~W.~Serene,
  J.~Low.~Temp.~Phys. {\bf 29}, 179 (1977).

\bibitem{Fejos:2011zq} 
  G.~Fej\H{o}s and Zs.~Sz{\'e}p,
  Phys.\ Rev.\ D {\bf 84}, 056001 (2011).

\bibitem{Marko:2013lxa} 
  G.~Mark{\'o}, U.~Reinosa and Zs.~Sz{\'e}p,
  Phys.\ Rev.\ D {\bf 87}, 105001 (2013).

\bibitem{Marko:2015gpa} 
  G.~Mark{\'o}, U.~Reinosa and Zs.~Sz{\'e}p,
  Phys.\ Rev.\ D {\bf 92}, 125035 (2015).

\bibitem{Helmboldt:2014iya} 
  A.~J.~Helmboldt, J.~M.~Pawlowski and N.~Strodthoff,
  Phys.\ Rev.\ D {\bf 91}, 054010 (2015).

\bibitem{Pilaftsis:2013xna}
  A.~Pilaftsis and D.~Teresi,
  Nucl.\ Phys.\ B {\bf 874} (2013), 594.
  
\bibitem{Marko:2016wtw} 
  G.~Mark{\'o}, U.~Reinosa and Zs.~Sz{\'e}p,
  Nucl.\ Phys.\ B {\bf 913}, 405 (2016).
  
\bibitem{Berges:2005hc} 
  J.~Berges, S.~Bors\'anyi, U.~Reinosa and J.~Serreau,
  Annals Phys.\  {\bf 320}, 344 (2005).

\bibitem{Patkos:2008ik} 
  A.~Patk{\'o}s and Zs.~Sz{\'e}p,
  Nucl.\ Phys.\ A {\bf 811}, 329 (2008).

\bibitem{Marko:2012wc}
  G.~Mark{\'o}, U.~Reinosa and Zs.~Sz{\'e}p,
  Phys.\ Rev.\ D {\bf 86} (2012) 085031.

\bibitem{Bjorken_Drell_RQF} 
  J.~D.~Bjorken and S.~D.~Drell,
  {\it Relativistic quantum fields}, McGraw-Hill Book Company, New York, 1965.

\bibitem{Reinosa:2011ut} 
  U.~Reinosa and Zs.~Sz{\'e}p,
  Phys.\ Rev.\ D {\bf 83}, 125026 (2011).

\bibitem{Bordag:2000tb} 
  M.~Bordag and V.~Skalozub,
  J.\ Phys.\ A {\bf 34}, 461 (2001).
  
\bibitem{Reinosa:2011cs} 
  U.~Reinosa and Zs.~Sz{\'e}p,
  Phys.\ Rev.\ D {\bf 85}, 045034 (2012).
  
\bibitem{Baker}
  G.~A.~Baker and J.~L.~Gammel,
  {\it The Pad{\'e} Approximant in Theoretical Physics},
  Academic Press, New York, 1970.

\bibitem{Bender}
  C.~M.~Bender and S.~A.~Orszag, 
  {\it Advanced Mathematical Methods for Scientists and Engineers I, Asymptotic Methods and Perturbation Theory},
  Springer New York, 1999.

\bibitem{Chiku:1998kd} 
  S.~Chiku and T.~Hatsuda,
  Phys.\ Rev.\ D {\bf 58}, 076001 (1998).
  
\bibitem{Hidaka:2003mm} 
  Y.~Hidaka, O.~Morimatsu, T.~Nishikawa and M.~Ohtani,
  Phys.\ Rev.\ D {\bf 68}, 111901 (2003).

\bibitem{Brown}
  L.~S.~Brown,
  {\it Quantum Field Theory}, Cambridge University Press, 1994.

\bibitem{Badalyan}
  A.~M.~Badalyan {\it et al.}, 
  Phys.\ Rept.\ {\bf 82} (1982) 31.
  
\bibitem{Hidaka:2002xv} 
  Y.~Hidaka, O.~Morimatsu and T.~Nishikawa,
  Phys.\ Rev.\ D {\bf 67}, 056004 (2003).
  
\bibitem{Patkos:2002xb} 
  A.~Patk{\'o}s, Zs.~Sz{\'e}p and P.~Sz{\'e}pfalusy,
  Phys.\ Lett.\ B {\bf 537}, 77 (2002).

\bibitem{Pilaftsis:2017enx} 
  A.~Pilaftsis and D.~Teresi,
  Nucl.\ Phys.\ B {\bf 920}, 298 (2017).
  
\bibitem{GSL}
  M.~Galasi {\it et al.},
  {\it GSL Reference Manual},
  http://www.gnu.org/software/gsl.
 
\bibitem{Shima}
  H.~Shima and T.~Nakayama,
  {\it Higher mathematics for physics and engineering},
  Springer-Verlag  Berlin Heidelberg, 2010.

\end{thebibliography}
\end{document}